\documentclass[prb,aps,twocolumn,superscriptaddress,showpacs]{revtex4}

\usepackage{graphicx}
\usepackage{color}

\begin{document}

\title{Dynamic Stripes and Resonance in the Superconducting and Normal Phases of YBa$_{2}$Cu$_{3}$O$_{6.5}$ Ortho-II Superconductor}
\author{C. Stock}
\affiliation{Department of Physics, University of Toronto, Ontario M5S 1A7, Canada}
\author{W. J. L. Buyers}
\affiliation{National Research Council, Chalk River, Ontario, K0J 1J0, Canada}
\affiliation{Canadian Institute of Advanced Research, Toronto, Ontario, M5G 1Z8, Canada}
\author{R. Liang}
\affiliation{Physics Department, University of British Columbia, Vancouver, B. C., V6T 2E7, Canada}
\affiliation{Canadian Institute of Advanced Research, Toronto, Ontario, M5G 1Z8, Canada}
\author{D. Peets}
\affiliation{Physics Department, University of British Columbia, Vancouver, B. C., V6T 2E7, Canada}
\author{Z. Tun}
\affiliation{National Research Council, Chalk River, Ontario, K0J 1J0, Canada}
\author{D. Bonn}
\affiliation{Physics Department, University of British Columbia, Vancouver, B. C., V6T 2E7, Canada}
\affiliation{Canadian Institute of Advanced Research, Toronto, Ontario, M5G 1Z8, Canada}
\author{W. N. Hardy}
\affiliation{Physics Department, University of British Columbia, Vancouver, B. C., V6T 2E7, Canada}
\affiliation{Canadian Institute of Advanced Research, Toronto, Ontario, M5G 1Z8, Canada}
\author{R. J. Birgeneau}
\affiliation{Department of Physics, University of Toronto, Ontario M5S 1A7, Canada}
\affiliation{Canadian Institute of Advanced Research, Toronto, Ontario, M5G 1Z8, Canada}

\date{\today}

\begin{abstract}

We describe the relation between spin fluctuations and
superconductivity in a highly-ordered sample of
YBa$_{2}$Cu$_{3}$O$_{6.5}$ using both polarized and unpolarized
neutron inelastic scattering.  The spin susceptibility in the
superconducting phase exhibits one-dimensional incommensurate
modulations at low-energies, consistent with hydrodynamic stripes.
With increasing energy the susceptibility curves upward to a
commensurate, intense, well-defined  and asymmetric resonance at
33 meV with a precipitous high-energy cutoff.  In the normal
phase, which we show is gapless, the resonance remains
surprisingly strong and persists clearly in Q scans and energy
scans.   Its similar asymmetric spectral form above T$_c$=59 K
suggests that incoherent superconducting pairing fluctuations are
present in the normal state. On cooling, the resonance and the
stripe modulations grow in well above T$_c$ below a temperature
that is comparable to the pseudogap temperature where suppression
occurs in local and low-momentum properties.  The spectral weight
that accrues to the resonance is largely acquired by transfer from
suppressed low-energy fluctuations. We find the resonance to be
isotropically polarized, consistent with a triplet carrying $\sim$
2.6 $\%$ of the total spectral weight of the Cu spins in the
planes.
\end{abstract}


\maketitle
\section {Introduction}

Spin fluctuations play a fundamental role in the superconductivity
of high-temperature superconductors. Intensive research has
spawned a plethora of theories for the interplay between
superconductivity and antiferromagnetism but no model is generally
accepted for the unusual behavior of the doped planar cuprate
superconductors. The search for new phases of matter, especially
in the underdoped region, as the boundary with the
antiferromagnetic phase is approached, continues to provide
stimulus for experiment and theory.~\cite{Kastner98:70} Many
recent theories for both the normal and superconducting states
have made specific predictions for the spin correlations and
therefore a detailed and complete study of $\chi''$($\bf{Q}$,
$\omega$) is essential to understanding the cuprates.

    One of the most interesting properties of the cuprate phase
diagram is the existence of a pseudogap phenomenon seen clearly in
tunneling, NMR, transport, and ARPES
experiments.~\cite{Timusk99:62}  Tunneling data have evinced clear
evidence of a gap-like structure in the density of states well
above the onset to superconductivity.~\cite{Renner98:80} Some of
the cleanest data have come from NMR studies which have found a
suppression of both the Knight shift and the relaxation rate
$1/T_{1}T$ in the normal
state.~\cite{Warren89:PRL62-1193,Yasuoka97:282}

    Many theories have been constructed to explain the
pseudogap.~\cite{Lee99:317}  Most recently an orbital current or
flux phase has been suggested which predicts circulating currents
flowing  in the copper-oxide
planes.~\cite{Wen96:76,Lee02:052,Chak01:15,Varma97:55} In the
static version of the theory~\cite{Chak01:15,Chak01:63} a hidden
order parameter, called the d-density wave (DDW), condenses at a
temperature well above the superconducting transition temperature.
Since, the static DDW order would produce a moment, this phase
should be directly observable with neutrons as new Bragg peaks,
and since orbital currents break an Ising-like symmetry, a gap
should exist in the spin excitation spectrum.  We have previously
found that no new Bragg peaks are observable in the normal phase
of the ortho-II ordered YBa$_{2}$Cu$_{3}$O$_{6.5}$ (YBCO$_{6.5}$),
thus excluding strong DDW order that breaks the symmetry at
$\bf{Q}$ = ($\pi$,$\pi$). We cannot exclude the possibility of
broken orbital symmetry at zero wave vector~\cite{Varma97:55} or
the possibility that the orbital currents are primarily
dynamic.~\cite{Lee02:052}

    Another theory to explain the variety of phases throughout the
cuprate phase diagram is the stripe theory.  Upon the introduction
of charge, this theory predicts the existence of several phases
whose structure is analogous to the nematic and smectic phases of
liquid crystals.~\cite{Kivelson98:393,Emery97:56,Kivelson02:0683}
In the stripe picture, charge and spins are spatially segregated
with the spins forming antiphase domains.  In the smectic phase,
long-range correlations exist producing incommensurate Bragg
peaks. In the nematic phase, the stripes only have orientational
order and therefore the antiferromagnetic correlations will be
short-ranged. Also, since the ground state of the stripe phase
breaks a continuous symmetry there should be no spin-gap. This
theory seems to give a good account of the phase diagram of the
La$_{2-x}$Sr$_{x}$CuO$_{4}$(LSCO or 214)~\cite{Wakimoto00:61}
systems which have been studied in great detail.  For YBCO$_{6.6}$
Mook \textit{et al.}~\cite{Mook00:404} obtained the first evidence
for stripes along the \textit{b} axis. The overall picture in the
YBa$_{2}$Cu$_{3}$O$_{6+x}$  system is still not clear, however,
since a complete study as a function of doping has not been
completed and, as we will also show, the structure of the spin
dynamics is highly dependent on structural disorder and
impurities.

    The most striking feature of the spin spectrum in the YBCO$_{6+x}$
system is the presence of an intense resonance peak at low
temperatures which is well defined in both momentum and
energy.~\cite{Mook93:70}  The resonance peak has been observed in
the YBa$_{2}$Cu$_{3}$O$_{6+x}$,
Bi$_{2}$Sr$_{2}$CaCu$_{2}$O$_{8+\delta}$ (BSCCO$_{8+\delta}$)
(Ref. \onlinecite{Keimer99:60}), and the single layer
Tl$_{2}$Ba$_{2}$CuO$_{6+\delta }$ (Ref. \onlinecite{He02:295})
systems suggesting that the resonance is a universal feature of
all cuprate superconductors. Many theories have been developed to
explain the resonance peak in terms of the spin-fermion and band
structure models with the Cu$^{2+}$ spins being coupled to the
quasiparticles.~\cite{Morr98:81, Norman:01:63, Liu95:75} Since
these theories predict that the resonance is a direct consequence
of a gap opening in the quasiparticle channel, no well-defined
resonance is predicted to exist in the normal phase.

\begin{figure}[t]
\includegraphics [width=80mm] {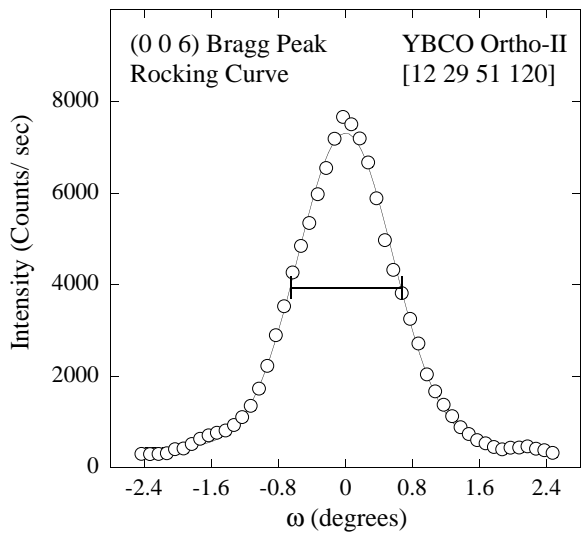}

\caption{Rocking curve of the
six-crystal composite sample at the (0 0 6) Bragg reflection. The
collimation from source to detector is given in minutes of arc within
square brackets.} \label{006rock}
\end{figure}

    We present a comprehensive study of spin fluctuations in the
oxygen ordered ortho-II YBCO superconductor in both the normal and
superconducting states up to $\sim$ 40 meV energy transfer. The
oxygen doping in the system studied YBCO$_{6.5}$
corresponds~\cite{TallonHoleDoping95:51} to a hole doping in each
CuO$_{2}$ plane of $p$=0.09.  We expect that the high degree of
structural order will remove some of the scattering and damping
seen in the spin response from samples where there is disorder
from twinning, poor chain order, and short oxygen chain segments
randomly located. We believe that much of the previous
discrepancies in the neutron scattering data can be reconciled by
the existence of such disorder and a consistent picture of the
underdoped region can be formed.

    This paper is divided into parts that deal with the normal
and superconducting phases.  The section on the normal phase will
present new results on the temperature dependence of the
low-energy excitations that show that no spin gap exists and that
one-dimensional incommensurate scattering is present that obeys
$\omega $/T scaling.  In the section on the superconducting phase
we show how the low energy scattering is suppressed, and that a
resonance peak grows that is much better defined in energy in YBCO
ortho-II than it is in disordered YBCO.  We end with a discussion
of the integrated intensities and the total moment sum rule, where
we determine the absolute weight of the resonance in the
superconducting phase and the surprisingly large fraction of
resonance weight that is already present in the normal phase.
Comparisons are made with the pseudogap phenomenon and with
theory.

\section{Experiment}

\begin{figure}[t]
\includegraphics[width=8.5cm]{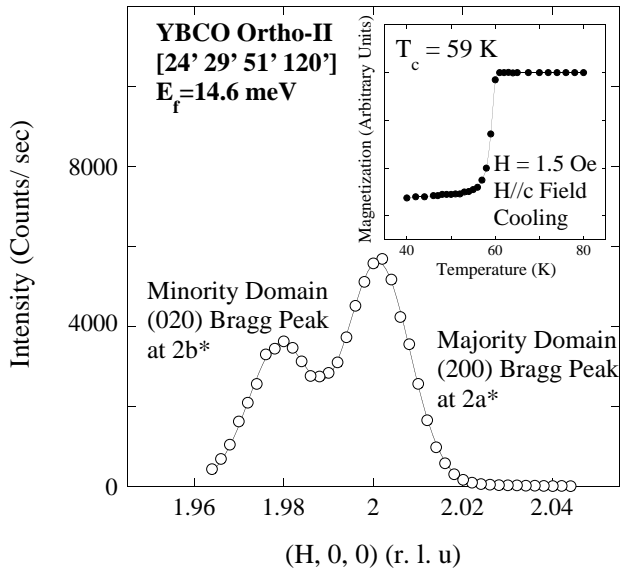}
\caption{Radial scan through the (2 0 0) Bragg peak.  The Gaussian fit
shows that the majority domain occupies $70\%$ of the sample volume.
The magnetization inset shows a sharp superconducting transition
temperature at 59 K with a width of $\sim$ 2.5 K.}
\label{200bragg}
\end{figure}

\begin{figure}[t]
\includegraphics[width=7.5cm]{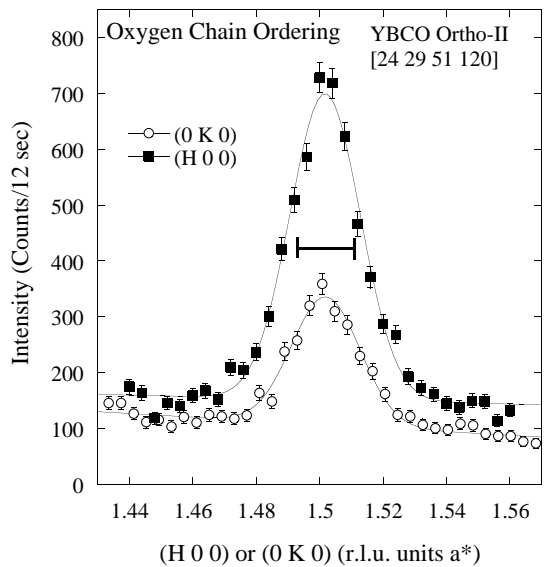}
\caption{Radial scans through the oxygen ordering superlattice
peaks at (3/2 0 0) and (0 3/2 0) (units of \textit{a*}).  The
horizontal bar indicates the resolution.  A fit to a
resolution-convolved Lorentzian shows the oxygen ordering
correlation length exceeds 100 \AA.  The larger peak (filled
squares) is from the majority domain and the lesser from the
minority domain.} \label{oxygen}
\end{figure}

    The sample consisted of six orthorhombic~\cite{Jorg90:41,Casalta96:258}
crystals of total volume $\sim$6 cm$^{3}$ aligned on a
multi-crystal mount with a combined rocking curve width of $\sim$
1.5$^{o}$ as shown in Fig. \ref{006rock}.  The lattice constants
were measured to be \textit{a}=3.81 \AA, \textit{b}=3.86 \AA, and
\textit{c}=11.67 \AA. Details of the crystal growth, the
detwinning by stress along the \textit{a} axis and the oxygen
order have been given earlier.~\cite{Stock02:3076, Peets02:xx} The
(2 0 0) scan of Fig. \ref{200bragg} shows that the majority domain
occupies 70$\%$ of the sample volume. The peak at larger
$|$\textbf{Q}$|$ (H=2) is the (2 0 0) Bragg peak from the majority
squeezed domain, and the peak H=1.98 is the (0 2 0) of the
minority domain. An independent check is obtained from the
satellites produced by oxygen chain order at (3/2 0 0) and (0 3/2
0). Fig. \ref{oxygen} shows scans through the oxygen superlattice
peaks.  Fits to resolution-convolved Lorentzians showed that the
oxygen correlation length exceeded $\sim$100 \AA~in the \textit{a}
and \textit{b} directions, while it was approximately 50 \AA~along
the \textit{c} direction.  The correlation lengths compare
favorably to the x-ray characterization of highly ordered ortho-II
indicating the high quality of our
crystals.~\cite{Andersen99:317,Liang00:336}

The measurements were carried out with the DUALSPEC triple axis
spectrometer at the C5 beam of the NRU reactor at Chalk River
Laboratories. A focusing graphite (002) monochromator and a
graphite (002) analyzer were used.  A pyrolytic graphite filter in
the scattered beam eliminated higher order reflections and the
final energy was fixed at E$_{f}$ = 14.6 meV.  For unpolarized
measurements the horizontal collimation was set at [33$'$ 29$'$
$S$ 51$'$ 120$'$] for energy transfer greater than 10 meV and set
to [33$'$ 48$'$ $S$ 51$'$ 120$'$] for energy transfers below 10
meV where the magnetic scattering is weaker. The vertical
collimation was kept fixed at [80$'$ 240$'$ $S$ 214$'$ 430$'$].
The six-crystal assembly was mounted in a closed-cycle
refrigerator that was carried on an open C-cradle. For
constant-energy scans along the in-plane directions the rotation
axis of the C cradle was along the [001] direction. This permitted
scans along (H H 0) in zones (1/2 1/2 L), and, by rotating the
C-cradle by $\sim$26$^{\circ}$, also along (H 3H 0) in the (1/2,
3/2, L) zone. To study the incommensurate nature of the magnetic
Q-correlated peaks, scans were made along the [H 0 0], [0 K 0],
and [0 0 L] directions independently with respect to the zone
center (1/2, 3/2, L).  To do this, the rotation axis of the
C-cradle was placed along the [$\overline{1}$10] direction. By
slaving the C-rotation to the L value we could access a general (H
K L).

    To confirm the magnetic origin of the scattering we made
polarized measurements with a Heusler (111) monochromator and
analyzer. A flipper coil was placed on the incident side and a
graphite filter was used on the scattered side to filter higher
order neutrons.   A Mezei flipper in the incident beam allowed
spin-flip (SF) and non-spin-flip (NSF) cross sections to be
measured. At the sample two pairs of coils applied either a weak
horizontal field (HF) parallel to $\bf{Q}$ or vertical field (VF)
perpendicular to $\bf{Q}$ (of $\sim$ 3-5 gauss) to control the
direction of the neutron spin at the sample. With the flipper on,
the difference between the HF and VF count rates gives the
spin-flip scattering from the magnetic electrons alone independent
of nuclear incoherent or phonon scattering.~\cite{Moon69:181}  To
prevent any possible depolarization from the Meissner state the
sample was always warmed to 100 K ($\sim$40 K above the transition
temperature) and then field cooled to low temperatures. During
scans the field direction with respect to the sample never changed
more than a few degrees thereby minimizing the effects of
depolarization on the neutron beam.

We measured the polarized scattering with two different Heusler
analyzers. The first small analyzer,  $7.9\times $3.2 cm$^{2}$ ,
was used for the constant energy scans.  The horizontal
collimation was set to [33$'$ 48$'$ $S$ 126$'$ 120$'$] and the
vertical collimation was [80$'$ 72$'$ $S$ 120$'$ 430$'$]. The
second taller analyzer of dimensions $13\times $12 cm$^{2}$ was
installed for the constant-{\bf{Q}} studies through the resonance.
The horizontal collimation was then set to [33$'$ 48$'$ $S$ 126$'$
120$'$] and the vertical collimation was [80$'$ 72$'$ $S$ 280$'$
430$'$]. The flipping ratios from both the small and large
analyzer crystals were measured at the (2 2 0) Bragg position for
E$_{i }$ = E$_{f }$ = 47.6 meV, the incident energy at the
resonance peak. The flipping ratio with the small analyzer was
found to be 19:1 for both VF and HF field directions.  The
flipping ratio for the tall analyzer was measured to be 15:1 for
the HF and 12:1 for the VF directions.  By using a taller analyzer
we were able to obtain a factor of two increase in the resonance
scattering per unit time.

    To compare our results with those of other groups and to
connect with theory we have put our measurements of
$\chi''$$(\bf{Q}, \omega)$ on an absolute scale.  We have
calibrated the spectrometer by measuring the integrated intensity
of an acoustic phonon near the strong (0 0 6) Bragg peak.  By
comparing the measured energy-integrated intensity to that
calculated in the long-wavelength limit, where eigenvectors are
known, we were able to obtain the calibration constant that puts
the magnetic intensities on an absolute scale. Details are
provided in the appendices.  The appendices also show how we made
the correction for monitor contamination by higher-order
wavelength neutrons.  This correction is particularly important at
small energies, where its neglect may lead to an underestimate of
the low energy response.

\begin{figure}[t]
\includegraphics[width=80mm]{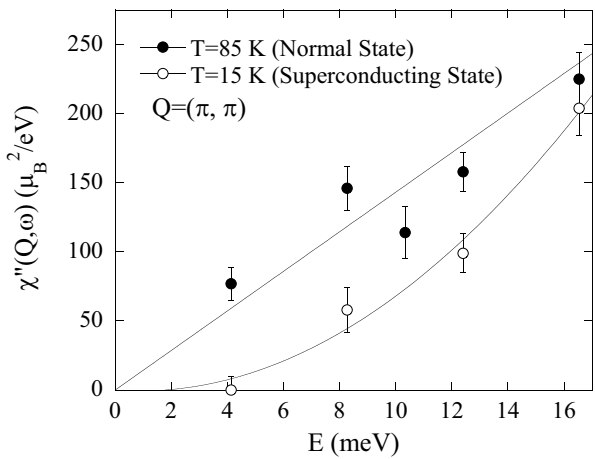}
\caption{The peak dynamic susceptibility, $\chi''$, at $\bf{Q}$ =
($\pi$,$\pi$) as a function of energy transfer for energies below
16 meV in the normal and superconducting states. In the normal
state scattering is present at the lowest energies studied and
increases roughly linearly with energy transfer (line). In the
superconducting state the scattering is suppressed but not
eliminated for energies below 16 meV (the lines are a guides to
the eye).} \label{chi_low_E}
\end{figure}

\section{The Normal State}
\subsection{Gapless normal state spin excitations}

    The existence or not of a normal state spin gap has many
important theoretical implications. Theories involving a broken
discrete symmetry such as orbital currents require the existence
of a normal state spin-gap at low temperature. In one model, the
d-density wave (DDW) theory, the orbital currents are equivalent
to static weak moments with Ising-like behavior, and the theory
predicts that a magnetic Bragg peak will appear in the underdoped
phase at temperatures well above the onset temperature of
superconductivity.~\cite{Chak01:15} In this model, in which the
orbital currents are in opposite directions in adjacent unit
cells, the DDW peak is predicted to break the spatial symmetry of
the copper oxide plane and to appear at $\bf Q$=$(\pi ,\pi)$.  In
the fluctuating orbital current theory of Wen and
Lee~\cite{Wen96:76,Lee02:052} there is no static order, and there
is a crossover to fluctuations that compete with superconducting
fluctuations in the normal phase. It is possible that this might
still lead to a suppression of the magnetic spectral weight well
above $T_{c}$.

    Experiments on underdoped YBCO in which no observable scattering
below $\sim$ 10 meV has been detected have been taken as evidence
for a pseudogap phase in the normal state.~\cite{Dai01:63}  In
terms of this interpretation the pseudogap is defined to be a
range in energy transfer where there is zero or negligible
spectral weight in the energy response as measured by the dynamic
susceptibility, $\chi''$($\omega$), and which lies below a range
with appreciable weight.  Another definition of the pseudogap
phase arises from NMR studies which have shown the
\textit{suppression}, not absence, of both the Knight shift and
the relaxation rate, $1/T_{1}T$, well above the superconducting
transition temperature.~\cite{Timusk99:62,
Hsueh97:56,Itoh96:65,Fujiyama97:66} Based on this we interpret the
normal phase pseudogap to be a spectral range in which either
there is negligible scattering or there is substantial suppression
of the scattering with decreasing temperature. Therefore, in terms
of neutron scattering, evidence for a pseudogap must come from
both the energy and temperature dependence of $\chi''$. Recent
work on near optimally doped LSCO has shown some evidence for a
spin pseudogap based on both the energy and temperature dependence
of the scattering.~\cite{Lee03:225} As regards the observed
dynamics we find no evidence for a gap or pseudogap in the normal
phase of YBCO$_{6.5}$ based on the temperature and energy
dependence of $\chi''$. In this section we present our results on
the energy dependence of $\chi''$ pointing to gapless excitations
in the normal phase.  The temperature dependence showing $\chi''$
increasing in the normal phase with decreasing temperature is
presented in later sections.

\begin{figure}[t]
\includegraphics [width=85mm] {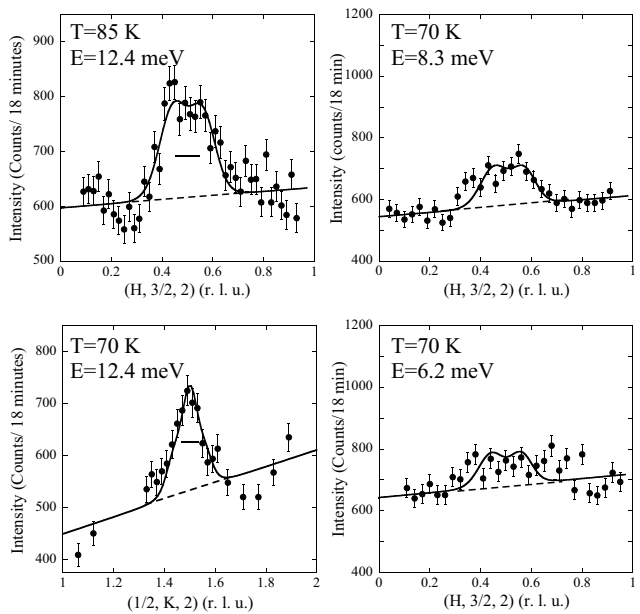}

\caption{Normal state neutron scattering scans through (1/2 3/2 2)
at energy transfers of 12.4, 8.3, and 6.2 meV.  The correction for
higher order contamination of the monitor rate has been made. In
the left two panels the scattering along [100] reveals
incommensurate features, but along [010] it does not (see text).
The solid curve is a fit of the spin-wave model of Chou \textit{et
al} discussed in the text. Their relative weight is given by the
70:30 domain ratio. The broken lines give the linear background. }
\label{chou}
\end{figure}

In an extensive survey of the low-energy spectra we have observed
that scattering centered on $\bf{Q}$ = ($\pi$, $\pi$) clearly
exists in the normal state and extends to the lowest energies
studied ($\sim$ 4 meV) .  A summary of our results, shown in Fig.
\ref{chi_low_E}, indicates the absence of any clear normal state
spin pseudogap in contrast to the suggested behavior by previous
studies.  The data were obtained by conducting fits to the
\textit{Q}-correlated peak above the background that lies under
the magnetic correlations centered on ($\pi$, $\pi$).  We have
removed the temperature-dependent Bose factor and the anisotropic
Cu$^{2+}$ form factor from the \textit{Q}-correlated scattering to
obtain the dynamic susceptibility as discussed in detail in the
appendices.

While the spectral weight at low energy is lower than that at the
resonance energy (see later), this, in itself, is not sufficient
evidence for a gap. We have recently discussed this result in the
context of orbital currents and have argued that the presence of
low energy excitations is evidence against the presence of a
static d-density wave or orbital current
phase.~\cite{Stock02:3076} From Fig. \ref{chi_low_E} we see that
the susceptibility decreases monotonically with energy below
$\sim$ 15 meV. The clear suppression of scattering in the
superconducting state will be discussed later. Clearly the
difficulty of detecting the normal-phase scattering increases at
low energies as the weight declines, but the trend is uniform and
gives no sign that the spectral response drops off in a way that
might suggest a gap. This linear decrease  of susceptibility with
energy transfer is consistent with over-damped spin waves or fermi
liquid behavior. It is also consistent with several early studies
of disordered YBCO in the underdoped region.~\cite{Chou91:43,
Sternlieb92:37,Birgeneau92:87}.  Note that the spin gap referred
to in several studies~\cite{Dai01:63,Tranquada92:46,Fong00:61}
(see  Figs. 9, 26 and 13 respectively) is the gap in the
superconducting state, not the normal phase gap whose absence we
address here.

Some typical scans at low energies along the [100] direction
through (1/2 3/2 2), are shown in Fig. \ref{chou}. This zone
boundary is equivalent to the 2D wave vector ($\pi$, $\pi$) with
L=2 near a maximum of the bilayer structure factor. The results
are corrected for the effect of higher order contamination of the
monitor rate. If this correction were ignored it would lead to an
underestimate of the low-energy intensity by a factor $\sim$ 1.5
at 6 meV as discussed in the appendices. The solid curves
represent the overdamped spin-wave model of Chou \textit{et
al.}~\cite{Chou91:43} (the following section) which gives an
excellent fit to the data and also predicts that the scattering
should increase roughly proportional to the energy transfer $\hbar
\omega$. Because it is more difficult to detect this signal as the
energy decreases, it is possible that experiments where the
sensitivity and statistics are optimized for the intense resonance
peak might have failed to observe the low-energy response and so
might have led to the inference that a gap was present. All of our
results indicate that there is no well defined normal state
spin-gap in Ortho-II ordered YBa$_{2}$Cu$_{3}$O$_{6.5}$.

\subsection{Incommensurate dynamic structure in YBa$_{2}$Cu$_{3}$O$_{6.5}$ Ortho-II}

To establish the incommensurate structure we made purely H and K
scans (see Fig. \ref{chou} left hand panels) in a configuration
with good in-plane resolution, where the coarse vertical
resolution was only 20-30$^{\circ}$ from the (0 0 L) direction and
so did not integrate over the incommensurate peaks.  We see that
the profile shows incommensurate peaks along \textit{a*} but only
a commensurate feature along \textit{b*}. From the H-K anisotropy
of the low-energy scans in Fig. \ref {chou} we can show that the
dynamic spin fluctuations are one-dimensional along \textit{a*}
with an incommensurate wave vector $\delta$ $\sim$ 0.06 r.l.u. To
do this we recognize that the scattering observed is the sum of
that from the majority and minority domains weighted in the
oxygen-order ratio of 70:30. Because of the low weight of the
incommensurate peaks of the \textit{b*} minority domain, the [010]
scan shows only a commensurate ridge as the scan passes between
the strong \textit{a*} peaks at ${\pm\delta}$ where the resolution
picks up its greatest contribution from the wings of the broadened
\textit{a*} peaks.

Since it gives a good description of the scan profile, we fit the
over-damped spin-wave model of Chou \textit{et
al.}~\cite{Chou91:43} and thereby extract $\delta$$\sim$0.06 after
convoluting with the spectrometer resolution function. In the Chou
model linear branches of damped spin waves rise isotropically in
the 2D plane from incommensurate wave vectors displaced by
($\delta$, 0) in r.l.u. from the antiferromagnetic point (1/2,
1/2), equivalent to $\bf{Q}$=($\pi$, $\pi$)~\cite{Q_notation}.
Similar behavior is found for low-energy excitations in the stripe
phase of the
nickelates~\cite{Tranquada95:375,Tranquada97:79,Bourges03:90,Boothroyd03:67}
where the stripes are static instead of fluctuating.  As we apply
the model to the \textit{a*}
 domain, $\bf{q}$ is measured from $\bf{Q}$ and the
incommensurate wave vector is $\bf{q}$$_{inc}$ = $(\delta , 0,0)$,
so that the Chou model is the sum of scattering from two wave
vector origins:

\begin{eqnarray}
\label{Eqn_Incomm}
{S(\bf{q}, \omega)|_{\bf{q}_{inc}}}&=& A {[n(\omega)+1]} {\hbar \omega \over
(\kappa^2+(\textbf{q} \pm \textbf{q}_{\textit{inc}})^2)}
\end{eqnarray}
\begin{eqnarray}
\times \left({\Gamma \over (\Gamma^{2}+(\hbar \omega - \hbar \omega_{\bf{q}\pm})^{2})}+
{\Gamma \over (\Gamma^2+(\hbar \omega + \hbar \omega_{\bf{q}\pm})^2)}  \right) \nonumber,
\end{eqnarray}

where

\begin{eqnarray}
{\hbar \omega_{\textbf{q}\pm}} & = & {\hbar c (\bf{q} \pm
\bf{q}_{\textit{inc}})}
\end{eqnarray}

and

\begin{eqnarray}
{\Gamma} & = & {\hbar c \kappa} = {\hbar c \over \xi_{\circ}}
\end{eqnarray}

\noindent where $c$ is the spin-wave velocity and $\xi_{\circ}$ can be
interpreted as a dynamic correlation length.  We emphasize that the parameter $\xi_{\circ}$ is
not strictly the instantaneous correlation length as we are not integrating over energy.
To account for
the minority domain we add to Eq. \ref{Eqn_Incomm} a similar term in
which the scattering emanates from wave vectors $\bf{q}$$_{inc}$ =
$(0, \delta,0)$.  Therefore the total cross-section is given as follows:

\begin{eqnarray}
{S(\bf{q}, \omega)} = {S(\bf{q}, \omega)|_{\bf{q}_{inc}=(\delta, 0, 0)}}+
{R S(\bf{q}, \omega)|_{\bf{q}_{inc}=(0, \delta, 0)}}.
\end{eqnarray}

\noindent The ratio of the two amplitudes, \textit{R},
 from the two domains is set to the known domain ratio of 30:70 determined
from the oxygen chain superlattice peaks.  After folding with the resolution
function the fit to the data is performed.

    The excellent agreement displayed in Fig. \ref{chou} shows that the
slow spin correlations are incommensurate in $\bf{Q}$. There is no
observable static incommensurate peak. At these low energies,
phonon contamination is unlikely and as a check we have verified
that the origin of this scattering at each energy transfer is
truly magnetic, from its temperature dependence around $T_{c}$
(see future sections regarding temperature dependence). By
iterating through each data set we have chosen the spin-wave
velocity, $\hbar c$ $\sim$ 300 meV$\cdot$\AA, a dynamic
correlation length, $\xi_{\circ}$$\sim$ 20 \AA, and
incommensurability, $\delta$=0.06, that applies at all energies
and temperatures.

    As shown in Fig. \ref{chou} the fit from this model describes the
line shape along both the [100] and [010] directions very well.
The minority domain is not strong enough to be seen as a separate
peak in the [010] scan (lower left panel), both because of its
small weight and because the majority peaks are broadened by $\xi$
and resolution so that they contribute most of the scattering  at
the commensurate point q=0. Thus the [010] scan shows a single
peak at K=1.5 as the scan crosses a saddle point between the two
majority peaks at ($\pm$ $\delta$, 0, 0).

    Our analysis provides strong evidence that the low-energy spin dynamics
reflect a one-dimensional incommensurate structure along
\textit{a*}. Its dynamics can be associated with the Doppler shift
for a right and left moving incommensurate spin density wave.
Similar incommensurate scattering has been previously observed for
oxygen concentrations of $x=0.6$~\cite{Mook00:404} and
$x=0.7$~\cite{Arai99:83} at energies of $\sim$ 20 meV, where the
incommensurate wave vector and doping is larger. The
one-dimensional incommensurate scattering is thus a common feature
of the YBCO$_{6+x}$ system in the underdoped region.
One-dimensional incommensurate scattering has also been observed
in La$_{1.95}$Sr$_{0.05}$CuO$_{4}$~\cite{Wakimoto00:61} indicating
that it is possibly a general feature of all cuprates.

\begin{figure}[t]
\includegraphics[width=80mm]{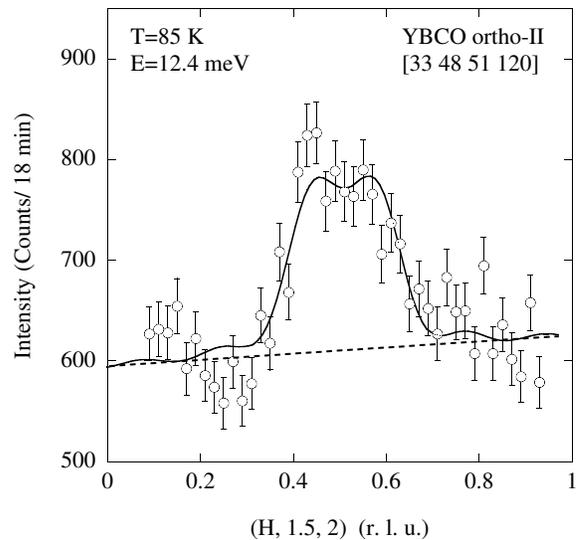}
\caption{The agreement of the stripe model (lines), in which
antiphase domains of length $\ell$=6 with $\delta$ = $1/2 \ell$ =
0.08 occur with a correlation length of 27 \AA, with the observed
scattering near the antiferromagnetic wave vector (1/2 3/2 2). }
\label{StripeFitNoMonCorr}
\end{figure}

    Physically, our data can be interpreted in terms of one-dimensional
antiphase domains parallel to the \textit{b*} chains.  This may be
consistent with the stripe picture~\cite{Tranquada97:166} in which
the spin correlations are antiferromagnetic in a domain between
charged walls separated by $\ell$ spins along [100], where the
carriers concentrate. Across a wall the coupling is ferromagnetic
so that the next spins belong to an antiphase spin domain $\pi$
out of phase with the first.  Antiferromagnetic coupling across a
row of spin vacancies would also give antiphase
domains.~\cite{Tranquada95:375} A nice feature of the stripe model
is that the local spin correlations within a domain remain
commensurate, as suggested by analysis of NMR
data.~\cite{Morr02:164} Nevertheless, the maximum in a scattering
experiment moves away from $\pi$ to an incommensurate wave vector,
$2\pi$/2$\ell$ i.e., $\delta$=1/2$\ell$, the inverse of the
2$\ell$ spin repeat distance for long-ranged stripe correlations.
Our data are consistent with the stripe model where the charge
domain is of length $\ell$=6 spins, and the domain correlations
fall off over a length of 7 spins ($\sim$27 \AA) as shown in Fig.
\ref{StripeFitNoMonCorr}. This fit was conducted by convolving the
stripe model with the $\bf{Q}$ resolution only and taking the
energy dependence to have a simple Lorentzian form. One of the
predictions of a sharp domain boundary is the presence of
higher-order satellite peaks which are clearly washed out by the
resolution.  This model gives a $\delta$ of 0.08, slightly higher
than that deduced using the Chou model but consistent given the
errors of the fit and the fact that we have neglected energy
dispersion in fitting the antiphase domain model.

    Because our data are consistent with the idea that an incommensurate
structure exists in the normal state, it is possible that the
stripe phase is a precursor to superconductivity in the underdoped
YBCO$_{6.5}$ system. The one-dimensional nature of the scattering
is difficult to interpret as originating from band structure
effects.~\cite{Kao00:61}  We find that for models such as Refs.
\onlinecite{Norman00:61} the anisotropic hopping caused by the
small 1$\%$ \textit{a-b} splitting should produce little
difference in the bands and peaks in ${\chi''}$ along the
\textit{a*} and \textit{b*} directions. This is confirmed by
density-functional calculations for the orthorhombic unit cell
which predict nearly isotropic hopping integrals in the
bandstructure.~\cite{Andersen94:49} It seems more likely that the
intrinsic structure of the spins (and carriers) in a plane favors
stripe order and that the small \textit{a-b} splitting as well as
the filled oxygen chains serves to select \textit{a*} to orient
the preexisting stripes.~\cite{Kivelson98:393, Emery97:56} Our
results are similar to those found in the LSCO system. In
particular, the incommensurate wave vector in this experiment is
consistent with those found in the monolayer cuprates with the
same hole doping.~\cite{Yamada98:57}

\begin{figure}[t]
\includegraphics[width=80mm]{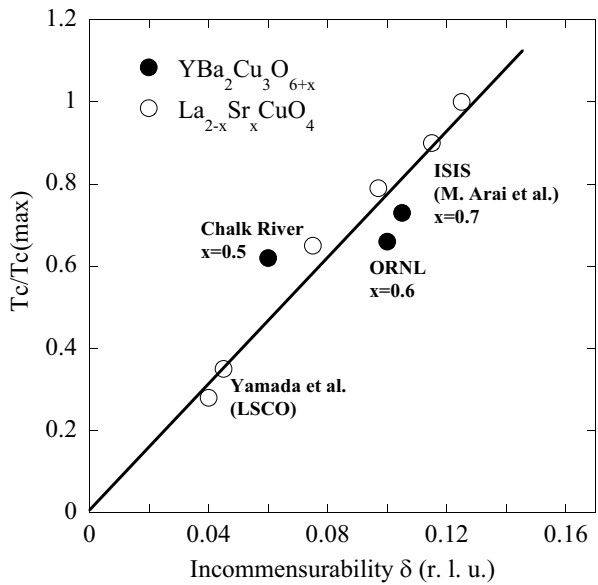}
\caption{The reduction of the superconducting transition
temperature upon reducing the oxygen doping below optimum, as it
relates to the position of the incommensurate peak in cuprate
superconductors. The incommensurate wave vector along the Cu-O-Cu
direction in reciprocal lattice units (r.l.u.) such that the wave
vector is $2\pi \delta /a$.}
\label{delta}
\end{figure}

    A comparison of the incommensurability with doping for bilayer and
single layer systems is shown in Fig. \ref{delta}.  As can be
seen, our data are consistent with values of $\bf{q}_{inc}$
observed in the LSCO system for similar hole doping. Even though
data in the YBCO system are currently very limited, the good
agreement to-date in the underdoped region points to a possible
universal trend between the hole doping and the spin stripe width.
Following Yamada \textit{et al.}~\cite{Yamada98:57}, Balatsky and
Bourges~\cite{BalatskyBourges99:82} showed that the overall
half-width at half maximum of the \textit{Q}-correlated peak below
the resonance, whether it shows incommensurate modulation or not,
is proportional to $T_{c}$. Our results are in agreement with this
relation. This suggests that the widths in YBCO in part arise from
incommensurate modulations and so can be interpreted within a
stripe picture similar to that in LSCO.

 The fact that $\delta$ is small, combined with short correlation lengths
and twinning, could explain why many studies in this doping region
have observed only a single commensurate peak at ($\pi$, $\pi$).
This point was the main assumption of the analysis of Chou
$\textit{et al.}$ who fit a series of commensurate inelastic
magnetic peaks to a model with spin waves originating, with equal
weight, from the two slightly incommensurate positions of
$\bf{q}$=$(0, \delta, 0)$ and $\bf{q}$=$(\delta, 0,
0)$.~\cite{Chou91:43}  It is also clear from our data that in a
sample with equally populated structural domains, a single
commensurate peak would be observed because of the small
incommensurability.

The incommensurate correlations along \textit{a*} are a robust
feature of the behavior and not a result of superconducting order
in underdoped YBCO. They are present in the normal phase up to
temperatures of at least $\sim$ 120 K, and, as we shall see later,
they persist in the superconducting phase up to $\sim$ 25 meV
(Fig. \ref{incom_6THz}).

In scanning tunnelling microscopy Davis, de Lozanne and
collaborators~\cite{DerroDavisYBCOPRL02:88} have observed a charge
modulation along the \textit{b*} (chain) direction in optimally
doped YBCO with a 13 \AA\  periodicity cell. This would correspond
to a reciprocal lattice wave vector of 0.3 r.l.u.  In underdoped
YBCO$_{6.5}$ there is no evidence for this charge modulation along
\textit{b*} nor of any double-length spin modulation 0.15 r.l.u.,
both of which would have been easy to resolve. The spin modulation
wave vector lies along \textit{a*}.  However, the scanning
tunnelling microscope images the chains not the electronic
structure of the planes.

\subsection{Temperature dependence of $\chi''$($\bf{Q}$, $\omega$)
and $\omega/T$ Scaling in the Normal State}

\begin{figure}[t]
\includegraphics[width=85mm]{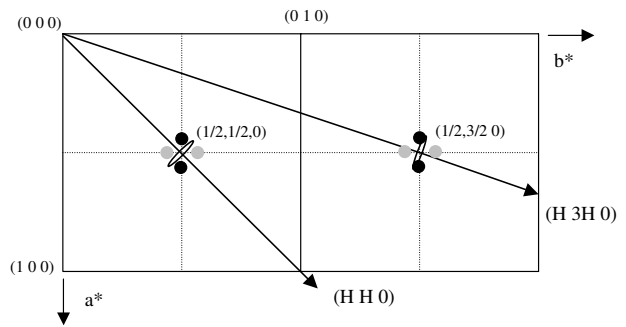}
\caption{ Illustration of in-plane wave vectors accessible in the
scattering planes (H 3H L) and (H H L).  The actual scans were
chosen to lie in the (H 3H L) plane so that the vertical
resolution (ellipse) would integrate over the stronger
incommensurate peaks along a*.} \label{H3H_plane}
\end{figure}

\begin{figure}[t]
\includegraphics[width=85mm]{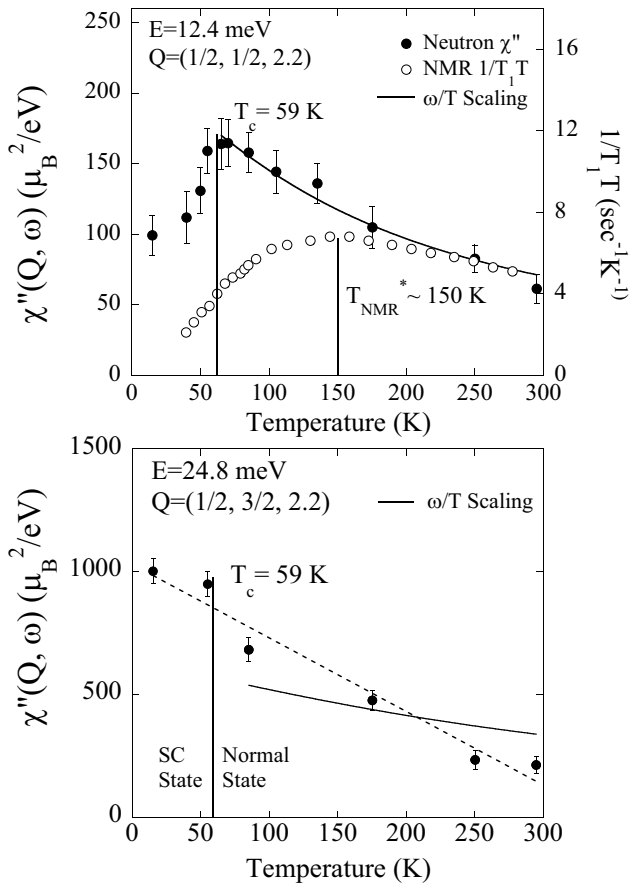}
\caption{  The peak susceptibility at 12.4 meV and 24.8 meV as a
function of temperature.  For the normal phase, the solid curves
are fits to the $\omega/T$ scaling analysis of Birgeneau
\textit{et al.} At 12.4 meV the normal state follows the scaling
analysis.  A clear suppression of scattering is observed in the
superconducting state. The scattering at 24.8 meV in both the
normal and superconducting states continues to grow on cooling and
no longer follows the nearly constant temperature dependence
predicted by $\omega/T$ scaling.  The dashed line in the lower
panel is a guide to the eye.  The local and low-frequency
susceptibility sensed by the NMR relaxation rate of $^{63}$Cu is
suppressed on cooling below a temperature T*, while the
susceptibility at the antiferromagnetic wave vector increases
uniformly until the onset of coherent superconductivity. NMR data
is for YBCO$_{6.64}$ taken from Timusk and Statt. }
\label{wT_compare}
\end{figure}

    To study the \textit{Q}-correlated peak as a function of energy and
temperature, the sample was aligned in the (H 3H L) and the (H H
L) scattering planes.  This configuration was chosen so that the
vertical resolution integrated over the intensity from the two
slightly out-of-plane incommensurate peaks (see Fig.
\ref{H3H_plane}). Since the width of the \textit{Q}-correlated
peaks was found, within experimental error, to be independent of
temperature, the peak susceptibility was then a good measure of
the integrated intensity.  To extract the susceptibility in
absolute units we have removed the anisotropic Cu$^{2+}$ form
factor multiplied by the bilayer structure factor as discussed in
the appendices.  The bilayer structure factor for odd-symmetry
fluctuations will be shown later to give a good account of the
L-dependence of the scattering.

    The temperature dependence of the susceptibility is shown in Fig.
\ref{wT_compare} for energies of 12.4 meV and 24.8 meV.  One of
the most striking features is the presence of a clear suppression
of the spin fluctuations at 12.4 meV below the superconducting
transition temperature, as will be discussed in the next section
devoted to the superconducting state.  In the normal state,
$\chi''$ increases with decreasing temperature.  As previously
discussed this is further evidence for the absence of a normal
state pseudogap in this energy range.    Most surprising is that
$\chi''$$(Q, \omega)$ continues to grow on cooling while the NMR
$^{63}$Cu relaxation $1/T_{1}T$ (proportional to
$\chi''$$(\omega)/\omega$) declines below about T$^{*}\sim$ 150
K.~\cite{Timusk99:62, Takigawa91:43} The T$^{*}$ from NMR has been
associated with a pseudogap, but it does not grow significantly
with underdoping (T$_{c}\leq$ 60 K),~\cite{Berthier97:282} as
would be expected for a pseudogap that many believe grows with
underdoping.   We cannot exclude a smaller gap than 4 meV (Fig.
\ref{chi_low_E}), but such a small gap would predict the maximum
NMR relaxation to occur at much too low a temperature $\sim$30 K
and cannot explain the observed T$^{*}\sim$ 150 K (estimate via
Eq. 21 of Tranquada \textit{et al.}~\cite{Tranquada92:46}).

\begin{figure}[t]
\includegraphics[width=80mm]{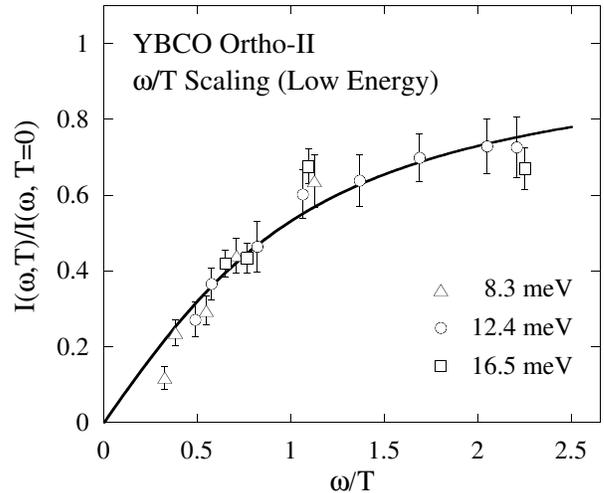}
\caption{A summary of the $\chi''$ is plotted for all normal state
data below 16 meV energy transfer as a function of $\omega/T$.
The temperature dependence of the low energy scattering is very
well described by $\omega/T$ scaling.} \label{wT_scale}
\end{figure}

    In early studies of the temperature dependence, $\chi''$ in
disordered YBCO$_{6.5}$ was found to follow a simple $\omega/T$
scaling relation.~\cite{Birgeneau92:87}  The same $\omega/T$
scaling was very successful in describing the temperature
dependence of the integrated susceptibility in underdoped
La$_{2-x}$Sr$_{x}$CuO$_{4}$.~\cite{Keimer91:67,Hiraka01:70}  The
analysis assumes that the dominant energy scale is set by the
temperature and predicts the following temperature and energy
dependence of the scattering,

\begin{eqnarray}
{\chi''(\omega, T) \over \chi''(\omega,T=0)}= {2 \over \pi}
\arctan \left({\omega \over a_{1} T}+{\omega^{3} \over a_{2}
T^{3}}+...\right).
\end{eqnarray}

\noindent Here the term $\chi''$$(\omega,T=0)$ represents the
limit of $\chi''$ as the temperature goes to zero.  A detailed
study on disordered YBCO$_{6.5}$ showed~\cite{Birgeneau92:87} that
the temperature dependence of all scattering up to $\sim$ 33 meV
is described well by just the first term with $a_{1}$ set to 0.9:

\begin{eqnarray}
{\chi''(\omega, T) \over \chi''(\omega,T=0)}= {2 \over \pi} \arctan \left({\omega \over 0.9 T}\right).
\end{eqnarray}

\noindent The inclusion of higher terms in the expansion was found
not to noticeably improve the fits. This scaling was also
originally proposed in the context of the marginal Fermi liquid
theory to explain many of the normal state physical properties of
the cuprates.~\cite{Varma89:63}  The marginal Fermi liquid theory
does not give the strong \textit{q}-space structure seen here.

    For ordered YBCO$_{6.5}$ in its ortho-II state we find that in
the normal phase this scaling form is in excellent accord with the
spin response at low-energies. Surprisingly the scaling function
fits the data with the \textit{same} coefficient $a_{1}$=0.9 as
for the disordered system. The relation is plotted in the normal
state for 12.4 meV and 24.8 meV transfers in Fig. \ref{wT_compare}
and shows excellent agreement at 12.4 meV but not at 24.8 meV. The
results for energy transfers below 20 meV are summarized in Fig.
\ref{wT_scale}. The figure was obtained by fitting (with
$a_{1}$=0.9) the $\omega/T$ scaling relation at each energy to
derive a single parameter $\chi''(\omega, T=0)$, from which
$\chi''(\omega, T) /\chi''(\omega, T=0)$ follows for all
temperatures and energies. We conclude that $\omega/T$ scaling
accurately describes the normal state low-energy $\hbar \omega$
and temperature dependence.

    Despite the fact that the scaling analysis works very well
for low energies, the scaling previously observed and predicted
suggests that for higher energy transfers the integrated
susceptibility should be only weakly dependent on temperature.
This is clearly not the case at 24.8 meV (Fig. \ref{wT_compare})
where the response grows strongly on cooling. It also does not
hold at higher energies where we see $\chi''$ increasing
continuously with decreasing temperature. This difference in
scaling in the normal state shows a clear departure of the
scattering in YBCO ortho-II from that in the disordered systems
previously studied which observed scaling up to at least 33 meV.
We note that a breakdown of $\omega/T$ scaling has been observed
at very low energies in La$_{1.98}$Sr$_{0.02}$CuO$_{4}$ and is
thought to originate from a small gap in the excitations due to
the out-of-plane anisotropy.~\cite{Masuda93:62}

    A possible explanation for the presence of one-dimensional
incommensurate scattering, $\omega/T$ scaling, and the breakdown
of the scaling relation at higher energies can be found in the
stripe model. The stripes consist of one-dimensional domains $\pi$
out of phase with each other.  As discussed by Zaanen \textit{et
al.}~\cite{Zaanen97:282} for a stripe liquid, where there is no
long-range order and stripes are fluctuating, there are two
extreme energy scales. The first is at high energy where the
domain walls appear static, and the second is the hydrodynamic
limit where the stripes are in a fluid state.  For the second
hydrodynamic (low energy) limit the energy scale of the system is
set by the temperature.  These assumptions underlying $\omega/T$
scaling suggest that the fluctuating stripe model provides a
natural explanation for the $\omega/T$ scaling we clearly observe
at low energies in the normal state.  It is also argued by Zaanen
\textit{et al.} based on hydrodynamics that $\chi''$ $\sim$
$\omega/T$ which can also be derived from the $\omega/T$ scaling
relation by expanding the expression to first order. Therefore
stripes do not predict the presence of a spin pseudogap in the
normal state but predict scattering at all energies and that
$\chi''$ $\sim$ $\omega$ at low energies. This is exactly the
model applied here to our low scattering and originally used by
Chou \textit{et al.}  This is further illustrated in Fig.
\ref{chi_low_E} which shows the scattering at very low energies
decreasing as $\omega$.

\begin{figure}[t]
\includegraphics[width=8.0cm]{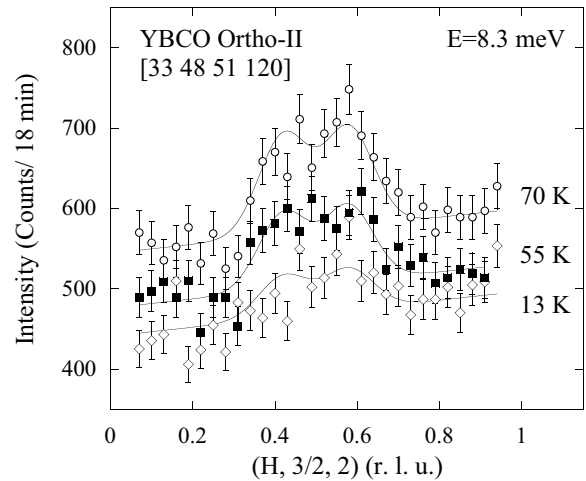}
\caption{Constant energy scans of the scattering at 8.3 meV energy
transfer. The incommensurate behavior persist unchanged apart from
amplitude upon entering in the superconducting state. The
background beyond the peaks and the fast neutron background scales
as [$n$($\omega$)+1].  The solid lines are from double Gaussian
fits to the data.}
\label{2THz_temp}
\end{figure}

    The stripe model also predicts the breakdown of $\omega/T$ scaling.  At
some point the energy will be sufficiently large compared with the
fluctuation frequencies that the fluid domains will appear almost
static. In our line-shape analysis this energy would be given by
$\omega_{c}$ $\sim$ $\hbar$c/$\xi_{\circ}$ = 300 meV $\cdot$\AA
/20 \AA = 15 meV. Therefore we would expect that for energies
above $\sim$ 15 meV the internal domain dynamics will start to
play a role in setting the energy scale and lead to a breakdown of
$\omega/T$ scaling.  This is qualitatively what is seen in our
data - perfect $\omega/T$ scaling up to at least 16 meV followed
by a clear departure from scaling at around 25 meV.

    These qualitative ideas can be extended to explain the role of
disorder. Disorder would decrease the correlation length $\xi$ and
therefore increase the characteristic frequency $\omega_{c}$. This
would provide a natural explanation for why early studies on
highly disordered crystals followed scaling up to high energies
$\sim$ 33 meV. These arguments provide some support to the idea
that the presence of fluctuating stripes in a liquid phase can
explain many qualitative features of the data in the normal state
including the temperature dependence and line shape.

    $\omega/T$ scaling can also be interpreted as the result of the close
proximity of a quantum critical point~\cite{Chubukov94:49} where
energy and temperature are interchangeable.  Such critical points
have been suggested to exist in the over-doped and near optimally
doped region of the generic cuprate phase
diagram.~\cite{Aeppli97:278,Varma99:83} The fact that we observe
scaling in the underdoped region and that scaling is observed over
a broad region in the LSCO system suggests that the scaling is not
the result of a quantum critical point. This assertion does depend
on the extent and size of the cross-over region.  To address the
issue of quantum criticality clearly a more detailed study as a
function of hole doping would be required in the YBCO$_{6+x}$
system.

\section{Superconducting State}

\subsection{Suppression of Low Energy ($\leq$ 16 meV) Scattering}

\begin{figure} [t]
\includegraphics[width=80mm]{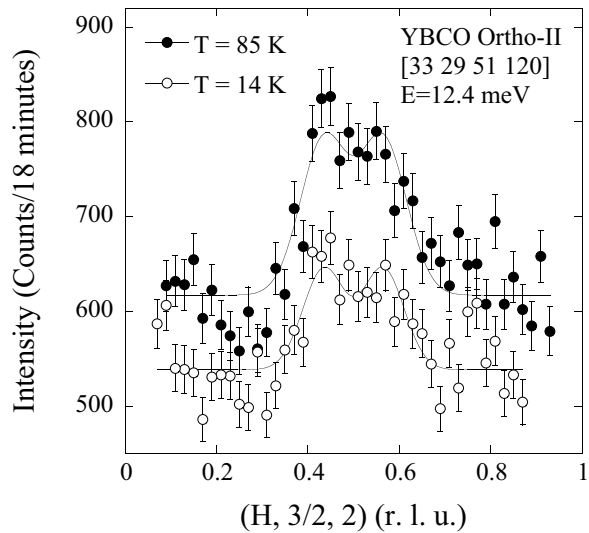}
\caption{Constant energy scans of the scattering at 12.4 meV
energy transfer. The line shape does not change upon entering in
the superconducting state.  The background scales as
[$n$($\omega$)+1].  The solid lines are from double Gaussian fits
to the data.} \label{3THz_temp}
\end{figure}

    On entering the superconducting state, the scattering at
 energies less than $\sim$ 16 meV is suppressed but not eliminated as shown
in Figs. \ref{chi_low_E}, \ref{wT_compare}, \ref{2THz_temp} and
\ref{3THz_temp}.  Thus there is not a complete gap.   Moreover,
the incommensurate structure remains in the superconducting state.
As shown in Fig. \ref{wT_compare} the suppression clearly starts
at $T_{c}$ and cannot be taken as evidence for a normal phase
pseudogap as previously discussed.  This suppression at the onset
of superconductivity clearly departs from the $\omega/T$ scaling
relation, in contrast to what was derived from early studies on
disordered systems for similar oxygen concentrations. Our
measurements also agree qualitatively with the result of early
studies on higher oxygen dopings which showed a departure from
$\omega/T$ scaling and a suppression of scattering in the
superconducting state.~\cite{Gehring91:44} This response is
consistent with the behavior in underdoped LSCO, where the low
energy scattering has been found to be suppressed in the
superconducting state~\cite{Yamada75:95} with almost complete
gapping at the lowest energy.~\cite{LakeNat400:99}  For
YBCO$_{6.5}$ we find that the energy below which the response is
suppressed in the superconducting state is consistent with the
\textit{gap} derived by other groups in the YBCO$_{6+x}$ system
for higher doping~\cite{Fong00:61} and with the universal curve of
Dai \textit{et al.}~\cite{Dai01:63} (see their Fig. 26).

     Thus, although there is no evidence for a normal phase gap,
there is clear evidence of formation of a gap as superconducting
order develops, with the degree of gapping increasing as the
energy tends to zero as seen in Fig. \ref{chi_low_E}. We do not
observe a full gap, and indeed would not expect to do so in a
superconductor where the pairing gap has nodes, regardless of the
gap symmetry.

    In contrast to the  suppression of the superconducting response below
$\sim$ 16 meV, for energies greater than 24 meV the response is
enhanced (Fig. \ref{wT_compare}).  It is interesting to compare
this observation to simple BCS theory which predicts a gap at
2$\Delta$ = 3.5k$_{B}$$T_{c}$ $\sim$ 18 meV.
Levin~\cite{Si93:47,Zha93:47}and coauthors have calculated
$\chi''$$({\bf{Q}}, \omega)$ in the random phase approximation
(RPA) and found a partial suppression of the response in the
superconducting phase. There is qualitative agreement with our
results (Figs. \ref{chi_low_E} and \ref{wT_compare}). The
temperature dependence of $\chi''$ for both the s-wave and d-wave
cases have been computed in the random-phase approximation by
Bulut and Scalapino.~\cite{Bulut93:47}  Their calculation was done
assuming a single band with only nearest neighbor hopping. The
hole doping was \textit{p}=0.15, somewhat above the doping of
\textit{p}=0.09 that is estimated for our ortho-II YBCO system
from its $T_{c}$. For a gap with s-wave symmetry the imaginary
part of the susceptibility shows a strong suppression or gap at
all energies up to $\hbar \omega \sim$ 2k$_{B}$$T_{c}$. The d-wave
model shows a suppression for energies below $\hbar \omega \sim$
k$_{B}$$T_{c}$ and an enhancement for energies of k$_{B}$$T_{c}$
and above. The d-wave results capture only the qualitative trend
of our data, in the sense that they show suppression at low
energies and enhancement at high energies. Quantitatively,
however, the model fails to predict the energy scale at which the
crossover from suppression to enhancement occurs, for it predicts
E $\sim$k$_{B}$$T_{c}$, which is $\sim$ 5 meV, whereas we observe
crossover at $\sim$ 16 meV or $\sim$3k$_{B}$$T_{c}$.  Despite this
disagreement the general trend of our data is consistent with the
formation of a d-wave instead of an s-wave gap which would show a
suppression of the scattering at all energy transfers.

\begin{figure}[t]
\includegraphics[width=80mm]{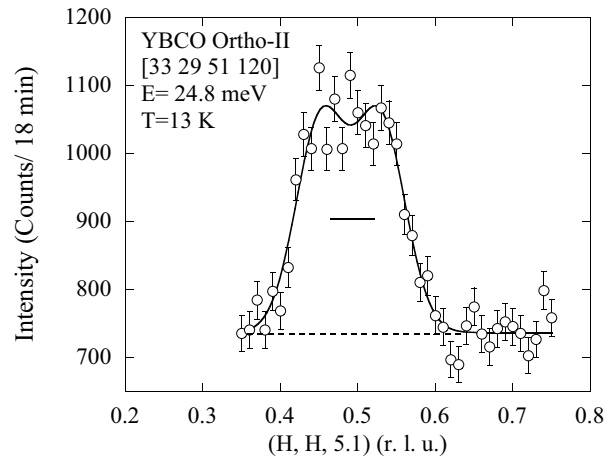}
\caption{Constant energy scans of the scattering at 24.8 meV
energy transfer showing a clear flat-top line shape indicative of
incommensurate scattering.  This scan was performed in the (HHL)
scattering plane.  The solid line is from a double Gaussian fit to
the data.} \label{incom_6THz}
\end{figure}

    As regards the spatial spin correlations we observe that they
continue to exist with the same incommensurate form in the
superconducting phase as they did in the normal phase, and this
despite the suppression of the susceptibility by the
superconducting order. The dynamic stripe structure is robust not
only with respect to temperature, but also with respect to energy,
for it is present up to $\sim$ 25 meV as illustrated in Fig.
\ref{incom_6THz}.

    In the context of the stripe picture previously discussed for the
normal phase, we observe that the spin suppression in the
superconducting phase occurs over an energy range similar to that
of the hydrodynamic region of the striped normal phase, i.e., the
energy range where we found that $\omega/T$ scaling describes the
normal state temperature dependence. This suggests that the
superconducting order acts to suppress the amplitude of the
precursor stripe fluctuations of the normal phase, but does not
essentially change their spatial or temporal character.

\subsection{Resonance at 33 meV}

\begin{figure} [t]
\includegraphics [width=85mm] {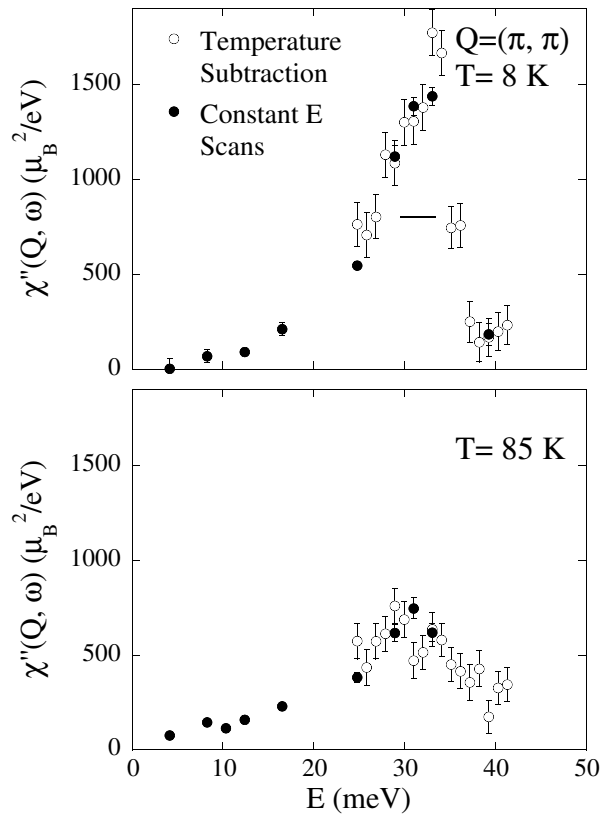}
\caption{$\chi''$ for all energies studied is summarized in both
the normal and superconducting states.  Two methods of measuring
$\chi''$ are plotted, the solid points being obtained from
constant-energy scans and the open points from a direct
subtraction of the high-temperature background susceptibility
measured at 250 K. A clear resonance appears in the
superconducting state and also in the data at 85 K, indicating
that a resonant feature does persist into the normal state.}
\label{chi_T_omega}
\end{figure}

\begin{figure}[t]
\includegraphics[width=80mm]{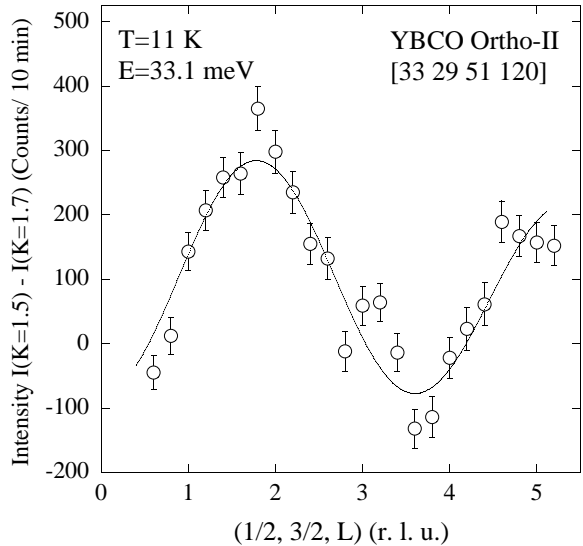}
\caption{Constant energy scan through the resonance along the [001]
direction.  Due to the presence of optic phonons a series of two point
scans were done with data being taken on and off the peak.  The solid
line is a fit to the bilayer structure factor.  The fact that the
subtraction gives negative values at the minimum of the bilayer
structure factor indicates the presence of a sloping background.}
\label{res_L}
\end{figure}

    One of the most dominant features of the scattering in
the superconducting state, as shown in Fig. \ref{chi_T_omega}, is
the presence of an intense and  well-defined peak in energy at 33
meV energy transfer.  We have verified that this resonance peak is
magnetic in origin from three features: it exhibits the clear
\textit{L} dependence (Fig. \ref{res_L}) expected for the bilayer
structure factor of the two copper oxide planes, its intensity
decreases with increasing temperature as does magnetic scattering
(Fig. \ref{int_res_E}), and it appears in polarized neutrons in
the purely magnetic difference channel, as will be discussed
later.

    In Fig. \ref{res_L} we show that the bilayer structure
factor, scaled by a single amplitude, gives an excellent account
of the modulation of the resonance intensity along the [001]
direction. The scan in Fig. \ref{res_L} was done by conducting a
series of one-point scans on and off the peak and subtracting. The
fact that the data are negative at the minimum of the bilayer
indicates the presence of a sloping background.  We have also
verified that the scattering follows the anisotropic Cu$^{2+}$
form factor by comparing constant energy scans done at the (1/2,
1/2, 2) position to equivalent scans conducted at higher
$|\bf{Q}|$ Brillouin zones.  This shows that the resonance is
consistent with scattering from Cu$^{2+}$ spins in the planes
which fluctuate in antiphase between the bilayers (also known as
the acoustic or odd-symmetry mode).

\begin{figure}[t]
\includegraphics[width=80mm]{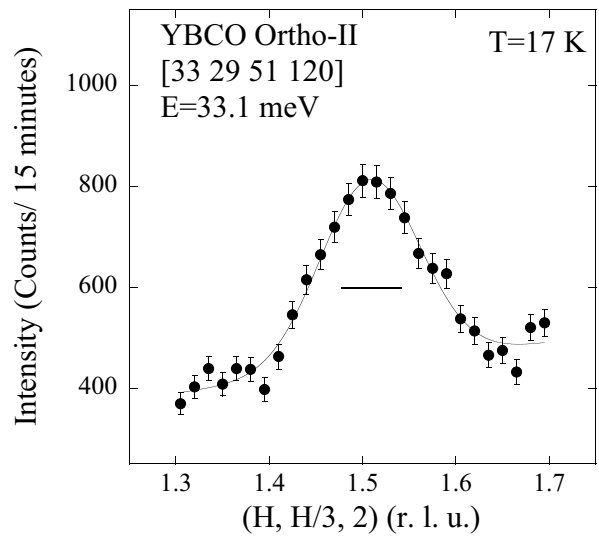}
\caption{Scan through the 33 meV resonance at (3/2 1/2 2) along (H
H/3 2), a direction that is sensitive to any incommensurate
modulation along \textit{a*} such as was observed well below the
resonance. A single commensurate resonance peak is observed whose
width exceeds the resolution (horizontal bar).}
\label{ResQscan3H_H_8K}
\end{figure}

    We find, as have others,~\cite{Dai01:63} that the resonance is
commensurate.  Scans along (K/3, K, 2), (H, H/3, 2), and (H, H, 5)
directions geometries have been used in previous studies to
investigate the incommensurate scattering as a function of energy
and temperature.~\cite{Bourges00:288}  We have determined that the
resonance is commensurate by scanning along the (K/3, K, 2), (H,
H/3, 2), and (H, H, 5) directions.  The latter two scans are
sensitive to any incommensurability in {\bf{Q}} along \textit{a*}
and show incommensurate peaks for energies below 24 meV. However
at the resonance we observe only a single commensurate peak, an
example of which we display in Fig. \ref{ResQscan3H_H_8K}.  Its
width (FWHM) in Q, after correcting for resolution, is $\sim$0.17
\AA$^{-1}$, implying a correlation length of 12 \AA\ or about 3
cells.

    In the ortho-II ordered sample the susceptibility shows a linear
region followed by a clear upward curvature to a well-defined
resonance at 33 meV (Fig. \ref{chi_T_omega}) with a half-width of
3.6 meV (including resolution). In contrast, when the system has
the same \textit{x}=0.5 oxygen content, but is disordered, the
linear region of the local susceptibility leads to a broad maximum
at $\sim$ 15 meV without the upward curvature typical of a
resonant mode.~\cite{Regnault94:235}  That broad spectrum extends
to low energies and differs substantially from the present
well-defined resonance of the well-ordered ortho-II system.  In
another \textit{x}=0.5 sample with T$_c$=52 K the peak of the
resonance occurred at $\sim$ 25 meV again without upward
curvature.~\cite{Bourges97:56} The large differences suggests that
the effect of structural disorder in the chains is to damp out the
resonance. Indeed the resonance is broadened in disordered systems
to give a tail on its high-energy wing, resulting in a more
symmetric peak, whereas in ordered ortho-II we find a sharp
resolution-limited decrease above the resonance. When the disorder
is small, as in the present ortho-II crystal, the resonance
becomes quite long-lived (of order 10 periods from the raw data
and longer if resolution is allowed for).  Also, this naturally
explains why in early studies the resonance was first discovered
in optimally doped systems. These have all the chains nearly full
and so no oxygen disorder is present (i.e., ortho-I chain
ordering), while for lower doping without stress detwinning a much
broader version of the resonance is
seen.~\cite{Regnault94:235,Bourges97:56} As for the effect of
disorder on the width in Q, we can understand why, in the
comprehensive study of Dai \textit{et al.}~\cite{Dai01:63}, the
sample with the most developed oxygen order, YBCO$_{6.6}$, as
determined from clear ortho-II oxygen peaks, exhibited a much
sharper resonance in $\bf{Q}$ than did other samples.

    This line of thought would suggest that the resonance should
be strongly broadened in the LSCO system because Sr enters the
lattice in a disordered manner, therefore having the analogous
effect to disordered oxygen chains in YBCO.  To-date there has
been no clear sign of a resonance in LSCO. However, there is
evidence for doubling of the strength of the local susceptibility,
relative to that in the antiferromagnetic insulator, at $\sim$20
meV, just above the superconducting gap.~\cite{Hayden96:76} This
could be taken as evidence that the superconducting order pushes
the spectral weight to a region above the
gap.~\cite{LakeNat400:99} If this feature was taken as the analog
of the YBCO resonance, its spectral form is substantially
different since it is broad with a tail to high energies, but may
nonetheless be consistent with the effects of disorder. The
resonance has been found in monolayer compounds indicating that a
bilayer structure is not necessary for its
presence.~\cite{He02:295}

\begin{figure}[t]
\includegraphics[width=85mm]{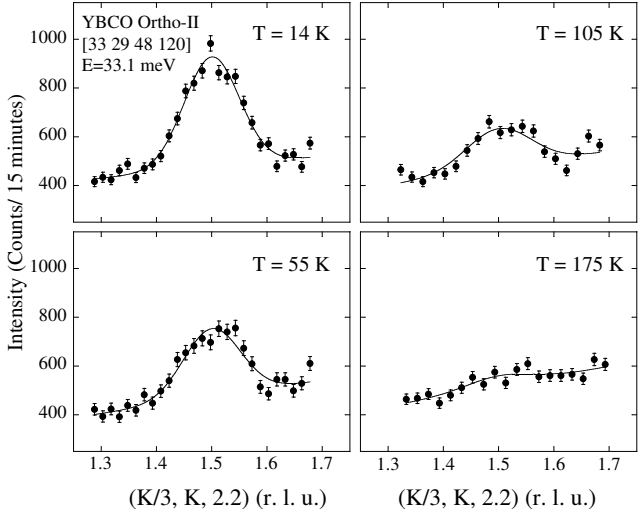}
\caption{Constant energy scans through the resonance peak as a
function of temperature.   The fits are to Gaussian peaks with the
fixed width for all temperatures. A clear peak is seen in the
superconducting state and it persists in the normal state until it
almost unobservable at 175 K.} \label{resonance_Q_temp}
\end{figure}

\begin{figure}[t]
\includegraphics[width=85mm]{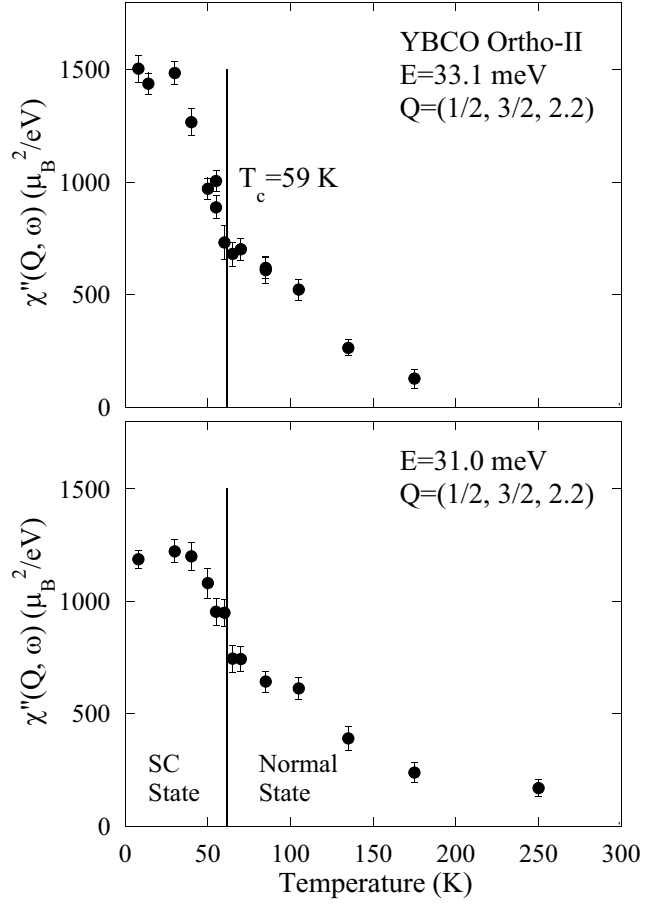}
\caption{The peak intensity as obtained from constant energy scans
of the scattering at 33 and 31 meV is plotted as a function of
temperature.  Both sets of data show a clear enhancement of the
intensity at $T_{c}$ and presence of a resonance peak at
$\bf{Q}$=(${\pi},{\pi})$ well into the normal state.}
\label{int_res_E}
\end{figure}

    Fig. \ref{resonance_Q_temp} shows constant energy scans at 33 meV
for several temperatures.  These scans show that the Q
correlations at the resonance energy still exist well into the
normal state.  Fig. \ref{int_res_E} shows the growth upon cooling
of the peak susceptibility at 33 and 31 meV. The spin response
exhibits a very clear upward break in slope at the onset of
superconductivity and is clearly enhanced below $T_{c}$. What is
plotted is the peak in absolute units derived from the
constant-energy scans of Fig. \ref{resonance_Q_temp}.

\begin{figure}[t]
\includegraphics[width=8.8cm]{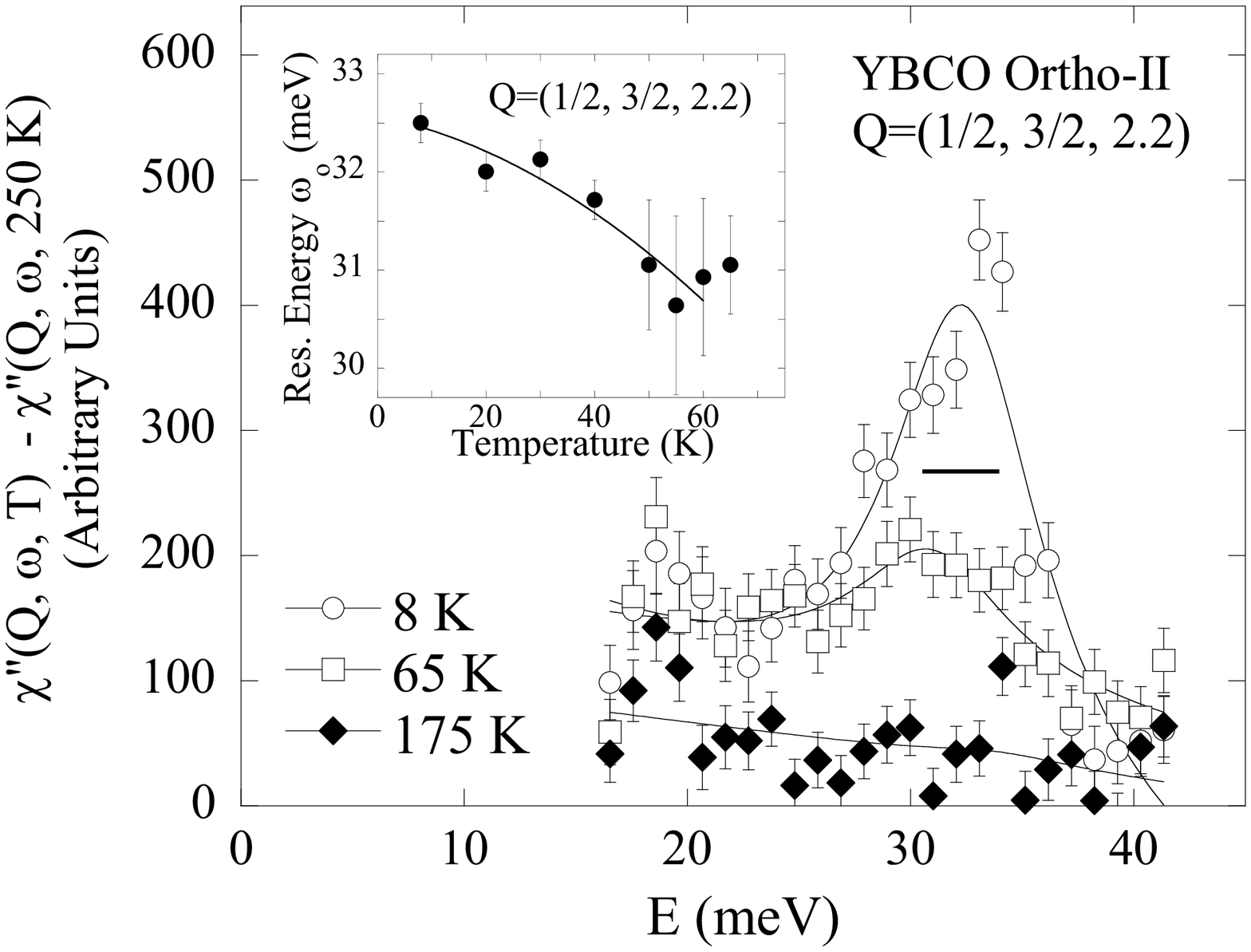}
\caption{Constant $\bf{Q}$ scans obtained from a temperature
subtraction with data at 250 K taken as the background and with
the Bose factor removed. The rise around 20 meV is due to a strong
optic phonon which does not completely subtract out in our
analysis, giving a sloping background. The data shows a clear
resonance in the superconducting state.  The 65 K data indicate
that a weakened resonance persists into the normal state. The
lines are a Lorentzian on a sloping background.  The inset shows
that the resonance energy decreases slightly with increasing
temperature.}
\label{res_omega}
\end{figure}

    To establish that in the normal phase the 33 meV feature is
resonant in time as well as localized in $\bf{Q}$ it is important
to observe that its width is less than its energy in a
constant-$\bf{Q}$ energy scan. To remove the effects of phonons we
have subtracted a background scan taken at 250 K. Since the
magnetic scattering, as determined from constant energy scans, at
energies greater than $\sim$ 24 meV, essentially disappears by 250
K this method of subtraction should give the full peak
susceptibility $\chi''$($\bf{Q}$, $\omega$).  We have also
subtracted an estimate for the temperature independent background
by rotating the analyzer crystal five degrees and counting.  As
can be seen from Fig. \ref{res_omega}, where results for three of
the temperatures studied are shown, the method of subtracting
high-temperature data does a reasonable job at removing the
phonons. There is a very intense optic phonon at 20 meV which does
not quite subtract out and gives an apparent rise in scattering
around this energy.   Nevertheless the energy scans show a clear
resonance below $T_{c}$ and reveal that in the normal state the
resonance remains as a weaker but still well-defined feature.

The lines in Fig. \ref{res_omega} are the results of a fit to a
Lorentzian and a temperature dependent background.  We fixed the
width at all temperatures to obtain a stable fit in the normal
state. The quality of the data above $T_{c}$ prevents us from
determining whether or not the resonance broadens in the normal
state.  The fit, though, does suggest that the peak of the
resonance shifts to higher energy (as shown in the inset to Fig.
\ref{res_omega}) with decreasing temperature, reminiscent of a
slight decrease in damping with decreasing temperature. This shift
in resonance frequency with decreasing temperature can be
explained by the spin-fermion model of Morr and
Pines.~\cite{Morr98:81} In the spin-fermion model the resonance
position is predicted to be inversely proportional to the magnetic
correlation length.  We should emphasize, however, that we do not
explicitly measure any temperature dependence of the correlation
length.  A similar result has been obtained by Fong \textit{et
al.}~\cite{Fong96:54} for optimally doped YBCO.

\begin{figure}[t]
\includegraphics[width=89mm]{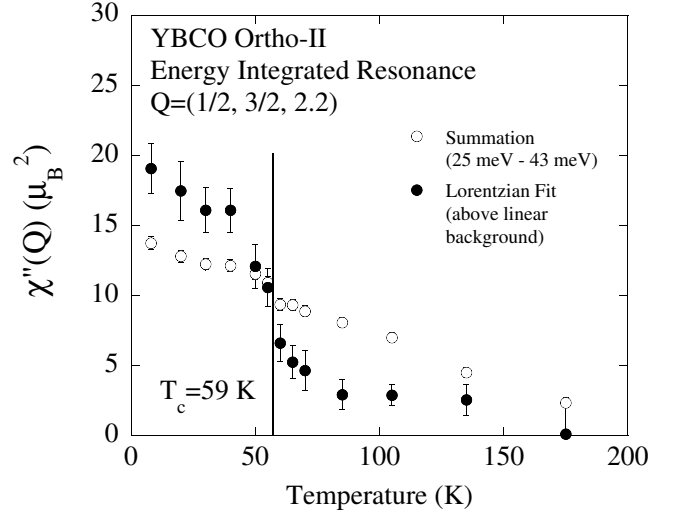}
\caption{The energy integrated intensity of the resonance as
measured in constant $\bf{Q}$ scans.  The integrated intensity in
absolute units was obtained by summing the data from 25 to 43 meV,
and also by fitting a Lorentzian to data like Fig.
\ref{res_omega}.  The latter method gives a resonance integral
that reaches a maximum of $\sim$ 19 $\mu_{B}^{2}$ at low
temperature. There is a clear increase in the growth rate of the
resonance intensity on entry to the superconducting state.  Both
methods show that a substantial fraction of resonant precursor is
already present in the normal state.}
\label{resonance_int_T}
\end{figure}

    The energy integrated  resonance intensity as a function of
temperature from constant $\bf{Q}$ scans is seen from Fig.
\ref{resonance_int_T} to show a clear enhancement below $T_{c}$.
Two methods were applied to estimate the energy integral. In one
(closed circles) we took the area under a fitted Lorentzian.  The
background fitted under the Lorentzian removes from the estimated
area part of the response that grows on cooling in the range 25-33
meV (in Fig. \ref{res_omega} compare T=65 K with 175 K).
Furthermore we know from low temperature data that the resonance
carries a long tail to low energies which is not accounted for by
a symmetric Lorentzian fit. Since we know the range 25-33 meV to
be free of phonons in our temperature-subtracted data (see
polarized data later), we adopted a second method for the spectral
weight in which we numerically summed the observed response from
25 to 43 meV (open circles in Fig. \ref{resonance_int_T}).  This
method will account, as we see, for less than the full integral at
low temperatures.

    The fraction of the full superconducting resonance intensity
that remains in the normal phase just above $T_{c}$ (Fig.
\ref{resonance_int_T}) is 25$\%$  from the Lorentzian fits and
70$\%$ from the numerical integration. Either estimate is a
substantial fraction of the low-T resonance. We conclude that the
resonant feature exists in the normal phase as a temporal and
spatially correlated feature, suggestive, as discussed later, that
superconducting fluctuations persist in the normal phase.  A
normal phase resonance at 34 meV was observed~\cite{Dai99:284}
below 150 K in YBCO$_{6.6}$ and used to define a pseudogap
temperature.  As a fingerprint for pairing it appears that of
order half the superconducting density, albeit incoherent, has
already formed in the normal phase.

    By combining both methods, $\bf{Q}$-scans and constant-$\bf{Q}$
energy scans using temperature subtraction, we have been able to
arrive at a complete picture of the magnetic spectrum up to 40 meV
as shown in Fig. \ref{chi_T_omega} in both the superconducting (8
K) and normal (85 K) states.  The spectrum is obtained at
high-resolution because we used a low final neutron energy of 14.6
meV for even the largest energy transfers.  The absolute
calibration was determined against the known cross-section of an
acoustic phonon assuming a paramagnetic (isotropic in spin)
cross-section (see appendices). The two methods for determining
$\chi''$($\bf{Q}$, $\omega$) agree very well, further indicating
the validity of the assumptions used in the subtraction analysis
previously discussed.

\begin{figure}[t]
\includegraphics [width=86mm] {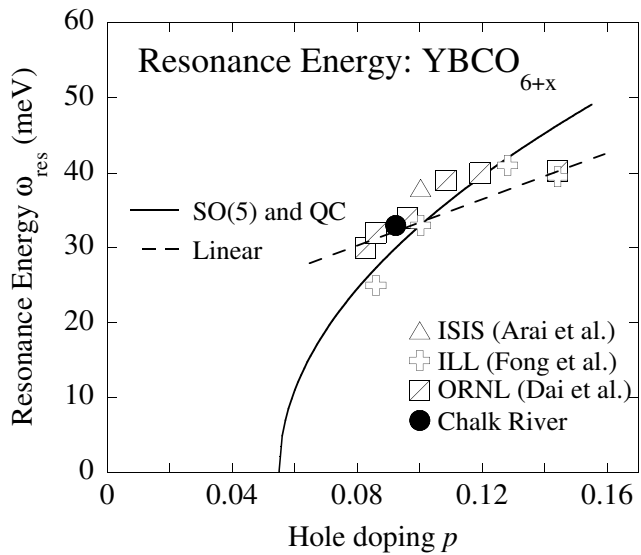}
\caption{A summary plot of the resonance energy as a function of
hole doping \textit{p} for the YBCO$_{6+x}$ system.  The solid
line is a fit $\omega_{res} \propto$ (p-p$_{c}$)$^{1/2}$ as
predicted by SO(5) theory.  A similar doping dependence for the
resonance energy was also predicted by Chubukov \textit{et al.}
based on the presence of a quantum critical point. The dashed line
is a linear fit. Both fits give reasonable descriptions of the
data.} \label{res_summary}
\end{figure}

    In Fig. \ref{res_summary} we plot the dependence of
the low temperature resonance frequency on doping
(derived~\cite{Loram97:282} from $T_{c}$) as determined by
different groups. It can be seen that the resonance frequency in
our sample is consistent with the overall trend. A linear scaling
of the resonance energy with $T_{c}$ (dashed line) is consistent
with most of the data. However the resonance trend also lies close
to the predictions of the SO(5)~\cite{Zhang97:275} and quantum
critical theory of Chubukov~\cite{Chubukov94:49} for which the
solid line is a fit to $\omega_{res} \propto$ (p-p$_{c}$)$^{1/2}$
with p$_{c}$=0.055, as measured carefully in the LSCO
system~\cite{Fujita02:65}. Because its spectral weight has a
strong onset at $T_{c}$ for optimally doped, and an accelerated
growth rate at $T_{c}$ for underdoped, there is little doubt that
the resonance peak is intimately related to superconductivity. It
is not yet clear which effect drives the other; the pairing could
involve the entire very high energy spin-wave band while the
resonance is a consequent low-energy concentration of spin
response caused by the pairing. Further studies, particularly at
lower doping, are required to address this point. We will discuss
this topic later in the context of the total integrated intensity
and the sum rule.

    At low temperatures we clearly see the presence of both a
commensurate resonance peak, and incommensurate scattering at low
energy transfers.  This coexistence points to a possible common
origin of both the resonance and incommensurate scattering.  This
topic has been discussed by Batista \textit{et al.} who have
analyzed possible common magnetic origins of both the resonance
and incommensurate scattering.~\cite{Batista00:30,Ortiz00:30}  In
the analysis of Batista \textit{et al.} the incommensurate
features are interpreted as spin-waves originating from static
incommensurate wave vectors.  A resonance then results when the
spin-wave branches meet at the ($\pi$, $\pi$) position.  This idea
connects with our data as our line shape found in the unpolarized
data has a clear tail at low energies and is suggestive of an
umbrella type dispersion where the excitations meet at the
resonance position. Such a dispersion in the context of the
incommensurate to commensurate behavior has been discussed in
studies of optimally doped YBCO.~\cite{Bourges00:288} However,
these ideas are speculative for YBCO$_{6+x}$ as the incommensurate
scattering is broad and we cannot resolve the spin-wave branches
suggested by this analysis.  Experiments on stripe-ordered
systems, however, give a spectrum with a sharp cut off for Q=$\pi$
at a maximum energy~\cite{Bourges03:90}, similar in form to the
asymmetric spectrum we observe for the YBCO resonance.


    It is not clear how the low-energy fluctuating stripes relate to
the resonance peak.  In the normal state we see formation of
incommensurate fluctuations below $\sim$120 K or 2$T_{c}$,
comparable to the temperature below which the resonance first
becomes clearly observable.   On cooling, the growth of resonance
intensity accelerates on passing into the superconducting phase,
concomitant with the onset of suppression of the stripe
fluctuations. Whether one or the other is the progenitor remains
an open issue.

\subsection{Polarization Analysis}

    To determine if there is any preferred orientation for the spin
fluctuations we studied the resonance and low-energy scattering in
three different zones and scattering planes.  These included the
(H, H, L), (H, 3H, L), and (3K, K, L) zones.  With unpolarized
neutrons the low-energy incommensurate scattering and the
resonance peak do not show measurable asymmetry between the (H,
3H, L) and (3H, H, L)  directions.  This rules out any enhanced
spin fluctuations oriented along either the \textit{a*} or
\textit{b*} directions. A comparison of the intensity in the (H,
H, L) plane with that in other scattering planes suggests no
preferred spin fluctuations along the \textit{c*} direction.  This
is the same isotropic spin polarization behavior as in a
paramagnetic phase.

\begin{figure}
\includegraphics[width=80mm]{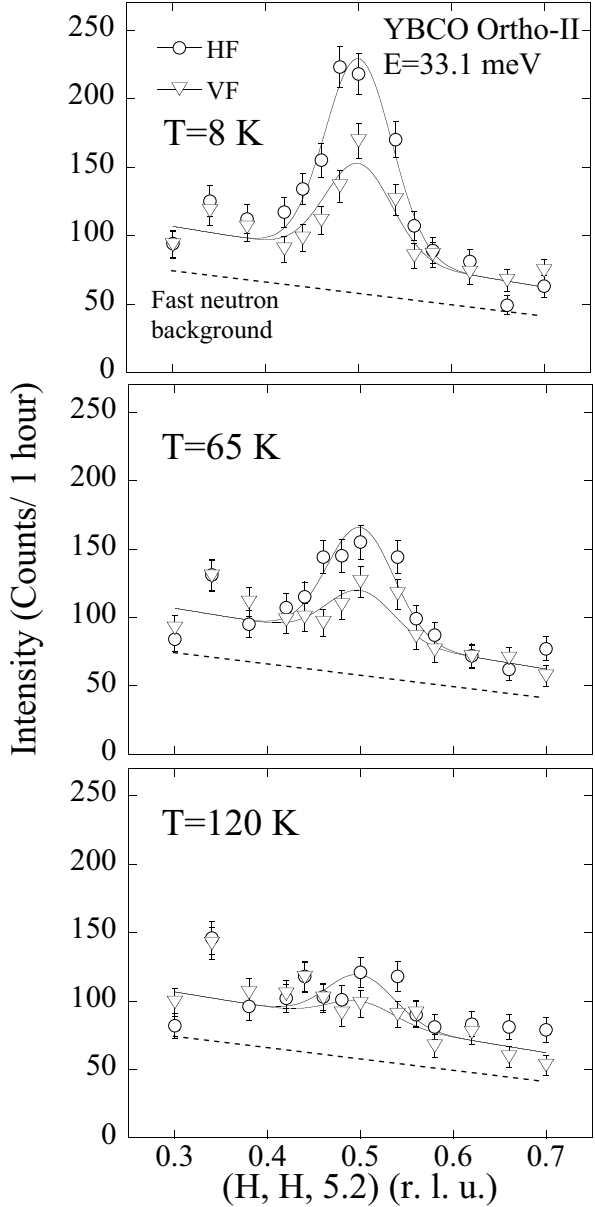}
\caption{Polarized constant energy scans through the resonance
peak at temperatures below and above $T_{c}$.  The decrease with
temperature of the resonance intensity at 8 K agrees with that of
the unpolarized data, eliminating any possibility of phonon
contamination. The open circles are taken with guide field
parallel to $\bf{Q}$ (HF) and the triangles with field
perpendicular to $\bf{Q}$(VF). The peak intensity in the VF
channel is roughly half that in the HF channel, which indicates no
preferred polarization of this excitation.  The dotted line is an
estimate of the fast neutron background. The feed-through from the
NSF channel can be estimated from the flipping ratio to be on
average $\sim$ 10 counts.  When added to the fast neutron
background, this almost completely accounts for the background
around the correlated peak.  The temperature dependent high point
at H=0.35 also appears in the NSF channel and is likely due to a
phonon. The solid lines describe Gaussian fits to the data.}
\label{res_q_polarised}
\end{figure}

    Further evidence for this finding comes from polarization analysis.  For
scattering from a paramagnetic material with a horizontal field
(HF) along {\bf{Q}}, all of the magnetic scattering will appear in
the spin flip channel since the neutron detects only the
perpendicular component of the magnetization. For a vertical field
(VF) perpendicular to {\bf{Q}} the magnetic scattering will
contribute equally to the spin-flip and non-spin-flip channels. If
the magnetic scattering is entirely paramagnetic, the scattering
in the HF channel will then be exactly twice that in the VF
channel. Fig. \ref{res_q_polarised} shows polarized scans through
the resonance at three different temperatures. The HF intensity is
always twice that in the VF channel to within error, showing that
the scattering arises only from the magnetic spins and that the
resonance is isotropically polarized. The fact that in this
paramagnetic system we see no preferred polarization in the spin
fluctuations can be taken to be consistent with transitions from a
singlet superconducting ground state to a triplet excited state.
We note that a triplet particle-particle resonance, coupled by the
superconducting order parameter into the particle-hole spin
susceptibility, is predicted by the SO(5) theory of Demler and
Zhang ~\cite{Demler98:396}.  We note that a triplet resonance was
an early prediction of Chubukov \textit{et
al.}~\cite{Chubukov94:49}

    The temperature dependence of the resonance intensity agrees
with that obtained from the unpolarized constant energy scans. The
agreement proves that magnetic resonance scattering persists in
the normal state at 65 K.  Since the scattering disappears at high
temperatures, this unambiguously demonstrates the absence of any
contamination from accidental Bragg or phonon scattering. This
type of contamination has been a problem in previous studies of
the resonance mode.~\cite{Mook93:70,Fong96:54}

\begin{figure}[t]
\includegraphics [width=85mm] {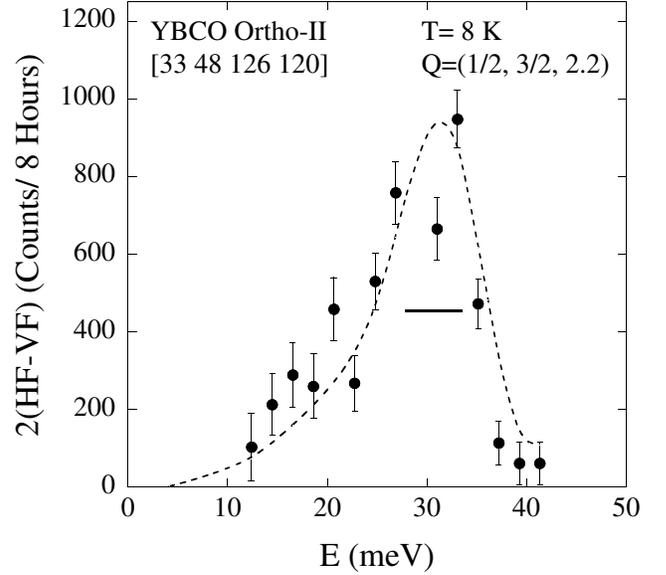}

\caption{Polarized measurement of the spectral form of resonance
peak obtained by subtracting the vertical field (VF) from the
horizontal field (HF) data .  The 33 meV peak position, the
resolution limited high-energy cut off, and the low-energy tail
connecting to the incommensurate response, confirm the spectral
form got from the unpolarized data (dotted line is smoothed
unpolarized data scaled by a constant factor).}
\label{polarised_E}
\end{figure}

    Fig. \ref{res_q_polarised} also shows that the background around
the resonance peak has the same intensity in the HF and VF
channels.  This indicates that there is no measurable continuum or
diffuse magnetic scattering other than the antiferromagnetic
correlations centered on the ($\pi$, $\pi$) position. This point
is further highlighted by the dotted line in Fig.
\ref{res_q_polarised} which gives the fast neutron count with the
analyzer rotated five degrees. Even though this method does not
obtain the full background, it does put a lower bound on the
amount of background scattering. From this one can see that most
of the background is taken into account by the scattering with the
analyzer turned. The rest of the spin-flip intensity in the wings
is close to that expected from feed-through from the non spin-flip
channel. This contribution to the scattering in the spin-flip
channel is equal to the scattering in the non-spin-flip (NSF)
channel divided by the flipping ratio. This is equal to
approximately 10 counts for the monitor used in Fig.
\ref{res_q_polarised}. The NSF feed-through and fast neutron rate
account for almost the entire background. This shows that there is
no measurable magnetic diffuse or continuum scattering around the
correlated peak at ($\pi$, $\pi$). Finally, we note that in this
energy range there is no observable scattering from hydrogen. This
would have occurred as nuclear spin-flip scattering equally in VF
and HF channels.  This attests to the careful mounting of the
sample in dry conditions~\cite{Stock02:3076} and to the minimal
amount of shielded adhesive used to seal the sample cans.

    Since it is established that the scattering is paramagnetic, we
have have been able to independently verify that the low
temperature spectral line shape derived from unpolarized data
(Fig. \ref{chi_T_omega}) is correct.  We subtracted the spin-flip
scattering at (1/2, 3/2, 2.2) in HF and VF channels and so
obtained the magnetic scattering alone, free from the phonon
contamination which we had to remove in analyzing the unpolarized
data. The resulting spectral form of the resonance shown in Fig.
\ref{polarised_E} is in excellent agreement with the unpolarized
results (the dotted line shows the smoothed unpolarized data). In
particular, the polarized scans confirm the interesting asymmetric
line shape consisting of a low-energy tail leading to a sharp
cut-off beyond the peak of the resonance as discussed above.

\subsection{Integral of S($\bf{Q}$,$\omega$) and the Sum Rule}

\begin{figure}[t]
\includegraphics [width=85mm] {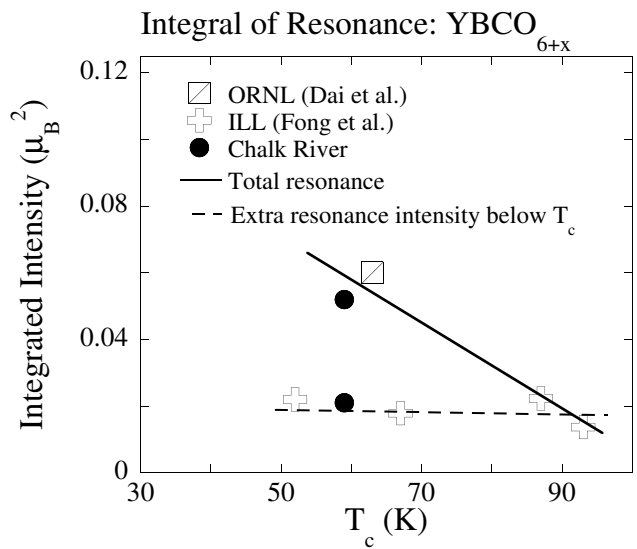}

\caption{The integrated resonance intensity, $\pi^{-1}$$\int
d\omega$$\int d^{3}q$ $[n(\omega)+1]$ $\chi''$($\bf{q}$,
$\omega$), measured by several groups for different values of
$T_{c}$.  An increase of spectral weight with decreasing doping is
predicted by both the spin-fermion and SO(5) models. The ILL data
has been divided by $\pi$ to correspond to the present definition
for the integral.  The upper Chalk River and ORNL points are the
total detected resonance weight.  The lower data are the growth in
resonance weight below $T_{c}$ for ILL and below 85 K for Chalk
River.} \label{integral_resonance}
\end{figure}

    From Fig. \ref{chi_T_omega}, we have estimated the total
integrated spectral weight of the resonance at 8 K. To do this a
symmetric Lorentzian profile centered at 33 meV was fit to the low
temperature energy scan in Fig. \ref{chi_T_omega} to obtain a
half-width $\Gamma$=3.6 meV and a peak intensity of 1400
$\mu_{B}^{2}$/eV. By assuming a Gaussian line shape along both the
[100] and [010] directions, a perfect column of scattering along
the [001] direction, and ignoring the bilayer structure factor, we
estimate the integrated scattering of the resonance to be

\begin{eqnarray}
I_{res}\equiv\pi^{-1} \int d\omega \int d^{3}q [n(\omega)+1]
\chi''({\bf{q}}, \omega)= \\
\nonumber 0.052 \mu_{B}^{2}
\end{eqnarray}

\noindent per formula unit where $d^{3}q$ is in reciprocal lattice
units. Our integrated intensity for the resonance compares very
well to the results of previous studies for similar oxygen
concentrations. Because of resolution effects, the true comparison
between different groups and experiments comes from the integral
of $\chi''$ in both momentum and energy. Fong \textit{et
al.}~\cite{Fong00:61} calibrated their measurements using an
acoustic phonon and have obtained an integral of 0.022
$\mu_{B}^{2}$ (0.069 $\mu_{B}^{2}/\pi$) for YBCO$_{6.5}$. As noted
in the appendices, the definition of $I_{res}$ used in that work
is $\pi$ times larger than the one used here. Dai \text{et
al.}~\cite{Dai99:284} have integrated the resonance to find $\sim$
0.06 $\mu_{B}^{2}$ for YBCO$_{6.6}$. This calibration was
conducted using the \textit{isotropic} Cu$^{2+}$ form factor
which, given the $\bf{Q}$ position in that experiment, would
introduce a small $\sim$ 5 $\%$ difference between our
measurements.  A summary of the absolute measurements of the
resonance as a function of $T_{c}$ is plotted in Fig.
\ref{integral_resonance}. The ILL and lower Chalk River data
points are the growth of the resonance integral that occurs below
$T_{c}$.  The ORNL and the upper Chalk River data points are the
total resonance integral.

    An increase in the spectral weight under the
resonance peak with decreasing doping is predicted by the SO(5)
theory of Demler and Zhang.~\cite{Demler98:396}  Their theory
relates the spectral weight in the superconducting state to the
difference in exchange energies between the normal and
superconducting states extrapolated to T=0, taken to be the
condensation energy, which is expected to decrease with lower
doping. As noted by Demler and Zhang this is consistent with the
electronic specific heat measurements of Loram \textit{et al.} who
have found a decrease in the condensation energy with decreased
doping.~\cite{Loram97:282} The condensation energy is proportional
to the difference in the integral of the dynamic susceptibility.
For the spin response below 42 meV we find that in
YBa$_{2}$Cu$_{3}$O$_{6.5}$ the integral $\int d\omega$
$\chi''$($\bf{q}$, $\omega$) (no Bose factor) is indeed larger in
the superconducting phase, since it amounts to 17 $\pm$ 2
$\mu_{B}^{2}$ at 8K and only 13 $\pm$ 2 $\mu_{B}^{2}$ at 85K, but
the difference lies within error.

    An increase with reduced doping of the resonance intensity
also arises in the spin-fermion model.~\cite{Morr98:81} With this
model, and previous normal state inelastic neutron results, Morr
and Pines were able to predict quantitatively a factor of $\sim$2
increase in the integrated intensity from optimal doping to
YBCO$_{6.5}$. The results of Fong \textit{et al}~\cite{Fong00:61}
show that any growth is by a smaller factor. However, Fong's
integral includes only the component of the integrated intensity
which develops below $T_{c}$.  We find the same integral if we
remove from our data all but the resonance growth below 85 K
(lower Chalk River point in Fig \ref{integral_resonance}).
However, our results for the total resonance weight of the normal
plus superconducting state, combined with the comparable results
of Dai $\textit{et al.}$, show that there is a clear growth of
total spectral weight with decreasing doping. These results lend
support to theories, such as SO(5) and the spin-fermion model,
which predict such an increase.

    Ignoring the effect of the chains, and therefore taking only
two Cu$^{2+}$ atoms per formula unit, the total spectral weight
integrated over all all energy and momentum should be given by
the total moment sum rule:

\begin{eqnarray}
\label{sum_rule_S} \int d\omega\int d^{3}q \ {S(\bf{q}, \omega)}=2
\times {2 \over 3} \textit{S (S+1)} g^{2} ,
\end{eqnarray}

\noindent  where $\int d^{3}q$ is the momentum space integral over
the correlated peak. In this equation the factor of 2 comes from
the fact that we are taking two Cu$^{2+}$ ions per formula unit.
Here we have assumed that $S(\bf{q}, \omega)$ has been corrected
for the bilayer structure factor and the Cu$^{2+}$ form factor
(see appendices).  We do not expect the total moment sum rule to
be strictly obeyed in the superconducting cuprates as the holes
will destroy a fraction of the Cu$^{2+}$ moments. Given that the
hole doping in our sample is \textit{p} $\sim$ 0.1, this effect
should be small and we expect the total measured moment to be
similar to that predicted by the sum rule. Equating the above
integral to the cross-section for paramagnetic scattering
discussed in the appendices, where the relation between $S(\bf{q},
\omega)$ and $\chi''(\bf{q}, \omega)$ is defined, we obtain the
total moment sum rule for localized spins as the integral over
$S(\bf{q}, \omega)$:

\begin{eqnarray}
\label{sum_rule_chi}
I \equiv \pi^{-1} \int d\omega \int d^{3}q  [n(\omega)+1]{\chi''(\bf{q}, \omega)}= \\
\nonumber {2 \over 3} \mu_{B}^{2} g^{2} S (S+1).
\end{eqnarray}

\noindent For S=$1 \over 2$ this gives a total integral of I$=$ $2
\mu_{B}^{2}$.  The total resonance at low temperatures therefore
makes up only $\sim$ 2.6 $\%$ of the spectral weight predicted by
the total moment sum rule. This agrees very well with estimates
found in other studies~\cite{Keimer99:60, Kee02:88} on
YBCO$_{6+x}$ and BSCCO$_{8+x}$. The absolute intensity of the
resonance has also been computed from band structure theories and
is in reasonably good agreement with our
experiment.~\cite{Norman00:61} Despite the fact that the resonance
is strongly affected by the approach and onset of
superconductivity, it contributes a tiny fraction of the total
moment. Its weakness has led to the suggestion that the presence
of a resonance is not crucial to the onset of superconductivity
and that it is unlikely the resonance can be associated with the
mediating boson of superconductivity.~\cite{Kee02:88, Abanov02:89}
This of course does not exclude the entire high-energy $\sim$250
meV spin-wave spectrum from providing the boson, while the
resonance is just a redistribution of a tiny fraction of the
low-energy spin spectrum in response to the opening of a pairing
gap, an idea that we now examine.

  To understand the origin of the resonance spectral weight we
compare in the superconducting and normal phases the integral from
0 to 42 meV of the two spectra of Fig. \ref{chi_T_omega}
multiplied by the Bose factor.  The integral, which is equal to
$\pi \int d\omega S(\bf{q}, \omega)$, was carried out by
polynomial interpolation of the data. If the resonance growth has
come from energies larger than our 42 meV window, or from other
momenta than than AF region, then we would calculate an increase
in the superconducting phase. For 8 K we obtained a $I=\int
d\omega \chi''(\bf{q}, \omega)$ $[n(\omega)+1]$ = 17 $\pm$ 2
$\mu_{B}^{2}$ and we find 14 $\pm$ 2 $\mu_{B}^{2}$ for 85 K.
Although there is some growth, within error the total weight of
the spin correlations is conserved. This indicates that the bulk
of the spectral weight gathering at the resonance energy at low
temperatures is transferred from from the low-energy
incommensurate scattering as this becomes suppressed in the
superconducting state. Since the Bose factor increases as $\sim$
$\omega^{-1}$ for small energy transfers, a natural concern with
this type of calculation is the weight carried by $\chi''$ at very
low energies, a region which is difficult to characterize
experimentally.  To assess its influence we note that that
integral from 0 to 5 meV at 85 K amounts to $\sim$ 0.5
$\mu_{B}^{2}$, representing only 4$\%$ of the total weight. Thus
the the low-energy limit of $\chi''$ is not contributing a large
amount and so validates the accuracy of the integral and our
finding of a low-energy to resonance transfer of spectral weight.

    We now extend our calculations to determine what fraction of
the total moment resides in the entire spectrum, not just the part
under the resonance.  By integrating numerically up to 42 meV, and
with the same q-dependence (Gaussian in the \textit{a-b} plane and
constant along the \textit{c} direction) we find a total integral
of $\pi^{-1} \int d\omega$$\int d^{3}q$
$[n(\omega)+1]$$\chi''$($\bf{q}$, $\omega$) $\sim$ 0.06
$\mu_{B}^{2}$.  We conclude that the spin response below 42 meV
carries about 3$\%$ of the spectral weight available from the
total moment sum rule.  Thus the spin response below 42 meV
captures only a small fraction of the total moment and illustrates
the importance of studying higher energy spin fluctuations.

\section{Discussion}

We first discuss the contribution to the absolute total moment
squared of the entire spectrum up to 43 meV. We have shown above
that this amounts to $\sim$ 0.06 $\mu_{B}^{2}$ per formula unit.
For a single Cu$^{2+}$ spin this gives $\sim$ 0.03 $\mu_{B}^{2}$,
comparable that observed to develop (on cooling in zero field)
(Ref. \onlinecite{Lake02:415}) in underdoped
La$_{1-x}$Sr$_{x}$CuO$_{4}$ with x=0.1, a doping similar to that
for the present YBCO$_{6.5}$. This is one of the many similarities
with the LSCO system.  We note that both LSCO and YBCO ortho-II
exhibit scattering that extends to the lowest energies in the
normal phase with incommensurate features characteristic of moving
stripes. There is a shorter correlation length in YBCO ortho-II,
yet we have shown that the incommensurate wave vector tracks the
doping equally in both systems (Fig. \ref{delta}).   For LSCO, if
we associate the resonance with the enhanced response at energies
above the region where superconductivity suppresses
$\chi''$($\bf{Q}$, $\omega$), then qualitatively there is much
similar behavior in LSCO and in YBCO.

    On cooling below $T_{c}$ the strong enhancement of spectral weight
in the resonance energy region is predicted by RPA calculations of
the susceptibility of a single quasiparticle band based on BCS
theory.~\cite{Bulut96:53, Norman00:61} The presence of resonance
intensity at temperatures above $T_{c}$ is not. We believe, as
discussed below, that this reflects local superconducting
fluctuations above $T_{c}$ in agreement with the evidence for
vortices seen through the Nernst effect.~\cite{WangOng02:88}

One aspect of the resonance observed here at high resolution
appears different from the results of previous studies of YBCO.
Its spectral shape is asymmetric with a sharp drop-off above the
peak, and with a much slower decrease along a wing below the
resonance. This wing connects smoothly with the previously noted
linear, low-energy response part of $\chi''$($\bf{Q}$, $\omega$).
Several calculations ~\cite{Norman00:61} predict the opposite
asymmetry as does the t-J
model~\cite{Fukuyama94:63,BrinkmannLee99:82}.
Norman~\cite{Norman00:61} has noted that the calculation for
$\chi''$($\bf{Q}$, $\omega$) drops off too fast below the
resonance. On the other hand the spin-fermion model of Morr and
Pines~\cite{Morr98:81} gives a spectral shape for the resonance
with a tail to low energies as we observe.

    The physical mechanism underlying the resonance peak still
remains unknown and is a matter of much debate.  Band structure
calculations using a single band associate the resonance peak with
the opening of a superconducting d-wave gap in the excitation
spectrum.~\cite{Norman:01:63,Bulut93:47} Since these theories are
mean-field they are not able to explain the presence of the
resonance in the normal phase at a similar energy to that in the
superconducting state.   Other theories such as that of Wen and
Lee, where d-symmetry normal-phase fluctuations are
dynamic,~\cite{Wen96:76} may be more appropriate. The QED$_{3}$
theory, which takes into account vortex-antivortex fluctuations
within the pseudogap region, may also account for the resonance we
see above $T_{c}$.~\cite{Franz02:66}

    A heuristic approach to understanding the spectral form of the
resonance is to interpret it as the result of a sampling of the
density of states of the d-superconducting (dSC) gap function. In
this picture the spectral weight of the overdamped low-lying spin
response is pushed up to energies above a varying gap function.
This accords with our observation of a transfer from low energies
to the resonance energy as superconducting order develops. We find
that a uniform sampling of a dSC form of gap function,
$\Delta_{\bf{k}}$=($\Delta_{0}/2$)($\cos(k_{x}$)-$\cos(k_{y}$))
gives a good account of the initial rise in $\chi''$, and also the
cusp-like cut off we observe at the maximum energy of the
resonance.~\cite{Buyers03:dSC} This might suggest that the spins
couple to an incoherent charge spectrum so that the momentum
conservation between the spin response and charge response of a
band model is compromised. It would imply that the cutoff beyond
the resonance might be a measure of twice the maximum gap,
2$\Delta _{0} $ $\sim$ 33 meV in a two particle picture. However,
the recent theory of Eschrig and Norman~\cite{EschrigNorman03:67},
which models the ARPES charge spectrum of
Bi$_{2}$Sr$_{2}$CaCu$_{2}$O$_{8+\delta}$, predicts for
\textit{p}=0.1 that the charge pairing gap $ 2\Delta _{M} $ = 120
meV $\sim$ 30k$_{B}$$T_{c}$ is much larger than our observed spin
resonance energy of $\sim$ 6$k_{B}$T$_{c} $, and that it increases
while the resonance energy declines with reduced doping. This much
larger energy scale presumably provides the less metallic
environment in which a spin resonance can survive with relatively
little damping.   An attractive feature of the
model,~\cite{EschrigNorman03:67} however, is that it predicts an
increasing degree of incoherence in the charged quasiparticle
properties as doping is reduced, a finding that lends support to
the momentum averaging that underlies our heuristic model.



We have looked for behavior that might account for the decline on
cooling of the NMR relaxation rate $1/T_{1}T$, which has been
associated with a pseudogap temperature of $\sim$150 K. We note
that the low-energy dynamic susceptibility in the meV range does
the opposite - it increases on cooling from room temperature right
through the pseudogap range, and continues to rise until
interrupted by superconducting pairing below 59 K.  Since the
normal response below 16 meV is linear in $\omega$, it follows
that its slope $\chi''$($\bf{Q}$, $\omega$)/$\omega$, which
amounts to 0.014 $\mu_{B}^{2}$/meV$^{2}$ and which is proportional
to the NMR relaxation rate 1/T$_{1}$T, also increases while the
NMR relaxation decreases. The different temperature dependence of
the NMR and neutron spin response, suggests that either the
extrapolation from the THz neutron to MHz NMR frequency range is
far from linear, or that it is at the other wave vectors sampled
by the local NMR probe that the pseudogap suppression takes place.

The only feature in the neutron spin response which has an onset
near the presumed pseudogap temperature is the resonance peak
which grows in below ~130 K. The increasing weight in the
resonance may be interpreted in terms of a shift of spectral
weight away from very low energies. In comparing the results in
the superconducting state with those in the normal phase, we have
noted earlier that the rapid increase of the resonance intensity
below T$_{c }$ can be accounted for by the loss of low-energy
intensity below $\hbar \omega$ $\sim$ 20 meV.  Such a picture
cannot account for the normal state, where the neutron spin
susceptibility inexorably rises on cooling and gathers extra
weight at the resonance frequency well above T$_{c }$ accompanied
by growth, rather than suppression at low energies. It is possible
that the spectral weight under the normal state resonance may be
acquired from higher energies or momenta, well removed from
($\pi$,$\pi$). However, since the resonance energy scale is
already much larger than thermal energies, it is unlikely that
there are substantial changes in the higher energy spectrum near
($\pi$,$\pi$). On the other hand, the depression of the local NMR
spin susceptibility (Fig. \ref{wT_compare}), which reflects an
average over momenta, suggests that much of the resonance weight
is being transferred from low momentum fluctuations.

While the spin correlations at all energies remain visible to the
highest temperatures as a correlated peak at ($\pi$,$\pi$), the
formation of the stripe modulation only becomes observable below
$\sim$ 120 K. This is in a broadly similar temperature range to
where we see the resonance begin to grow in. The emergence of
resonance intensity and stripe modulations are associated in the
model of Batista \textit{et al.}~\cite{Batista00:30}  We emphasize
that Batista's theory is for an incommensurate static ordered
structure, which is far from the present situation.

    Calculations for the 2D Hubbard cluster
model~\cite{MaierJarrell02:08419} have led to the conjecture that
uniform rather than antiferromagnetic spin fluctuations are
causing the pseudogap. However, they find no evidence for pairing
fluctuations for T$_c$$<$T$<$T*, and reject the notion of
preformed pairs in favor of RVB spin-charge separation.  Many of
the problems of the single-band theories may indeed be overcome by
a theory (such as spin-charge separation) where the dynamics are
not described in terms of individual
quasiparticles.~\cite{Anderson00:341} Spin-charge separation is
also favored in the theory of P. Lee~\cite{Lee99:317}.

    In the energy spectrum the pseudogap is clearly seen as noted
above in the charge response given by the optical
conductivity.~\cite{Timusk99:62,Homes93:PRL71} For YBCO$_{6.7}$
the c-axis conductivity is suppressed below an energy of $\sim$ 50
meV.  It grows in at high temperatures $\sim$ 3T${_c}$. For
YBCO$_{6.6}$ the suppression of the a-b optical conductivity is
below 80 meV.~\cite{Timusk99:62}  The connection between the
resonance in the spin susceptibility and the charge dynamics has
been made in the work of Schachinger \textit{et al.} and Carbotte
\textit{et al.}.~\cite{Schachinger01:54,Carbotte99:401}

    The resonance has been described as a two-particle
state~\cite{Demler98:396} or as a spin
exciton.~\cite{TchernyshyovNormChub01:63} Demler and
Zhang~\cite{Demler98:396} have described in the SO(5) theory how a
particle-particle (p-p) triplet resonance arises from paired
carriers.  The spin resonance at 33 meV then gives the maximum
energy, 2$\Delta$$_{0}$ to create a pair when the d-wave
superconducting (dSC) gap function is
$\Delta$=($\Delta_{0}/2$)(cos(k${_x}$)-cos(k$_{y}$)). The pair
resonance can only be seen in neutron scattering when it is
coupled to the spin susceptibility by the superconducting order
parameter and hence not above $T_{c}$. A reduction of the
superconducting order by a magnetic field would then lessen the
coupling to the spin channel as found by Dai \textit{et al.}, who
observed the resonance intensity to weaken with
field.~\cite{Dai00:406} However Tchernyshyov \textit{et
al.}~\cite{TchernyshyovNormChub01:63} have shown that for flat
bands near $(\pi,0)$ the neutron resonance is more likely a spin
exciton near the lower energy of $\Delta$$_{0}$.  The triplet
symmetry of the spin resonance was predicted by the quantum
critical theory of Chubukov \textit{et al.}~\cite{Chubukov94:49},
and also in SO(5), and is confirmed by our polarized neutron
results which can be interpreted in terms of a singlet ground
state with a triplet excited state.

It is more difficult to understand the presence of the spin
resonance in the normal phase, albeit weakened but of similar
spectral form. Indeed, most theories predict that the resonance
should be seen in the superconducting but not in the normal phase
above T${_c}$. Our observation of the resonance above T${_c}$ does
not support these models.  The quantum critical spin-fermion model
of Abanov \textit{et al.}~\cite{AbanovChubSchmalian03:52} (Eq. 140
with $\omega_{sf}$=33 meV), including hot spots, does give a broad
peak in the normal phase spin response. It is a relatively
featureless Lorentzian spectrum, however, falling slowly above a
spin-fluctuation energy.  The model does not predict the more
peaked spectrum with a faster high-energy fall off near the
resonance energy that we see in the normal phase (Fig.
\ref{chi_T_omega} lower panel). The spectral shape of the
normal-phase spectrum appears more like a weakened version of the
superconducting resonance spectrum and provides circumstantial
evidence that local superconducting fluctuations exist above
T${_c}$. Thus the neutron response supports the evidence that
fluctuating vortices exist in a YBCO$_{6.5}$ sample (T${_c}$=50 K)
as has been obtained from the Nernst effect.~\cite{WangOng02:88}
It is possible, we suggest, that the normal phase fluctuations may
be of the correct symmetry to let a local resonance appear in the
neutron spin response.  Since the neutron spectrum is independent
of the sign of $\Delta$$_{0}$, fluctuations of the gap parameter
on a timescale slower than the inverse resonance energy should
allow a resonance feature to remain in the normal state. The
density of states of the dSC gap function is strongly peaked at
its maximum energy, and this characteristic energy can persist in
the normal phase.  The loss of pair coherence on heating through
the dSC superconducting transition can occur in the nodal regions,
without eliminating the high-energy resonance spectrum. By
contrast in an s-wave superconductor the nodeless order parameter
is non-zero everywhere, and the gap function would have to
collapse in all directions to effect the transition. Extensions to
the theoretical models to treat fluctuations so as to extend their
validity to the normal phase would be valuable.

\section{Conclusion}

We have shown that the antiferromagnetic spin fluctuations in
YBCO$_{6.5}$ ortho-II are gapless, with incommensurate modulation
at low energy consistent with dynamic stripes.  The spin
susceptibility curves upward with energy to a well-defined but
asymmetric resonance peak at 33 meV followed by a precipitous
cut-off.  A weakened image of the resonance is readily observed in
the normal phase as a well-defined temporal and spatial structure
with an asymmetric spectrum similar to that of the superconducting
state.  On cooling, the resonance at 33 meV ($\sim$6.5
k$_{B}$$T_{c}$) and low-energy spin fluctuations strengthen until
the superconducting phase at 59 K is reached.  As coherent
superconducting order is established the resonance accelerates its
growth rate, while the low-energy fluctuations below $T_{c}$, are
sharply suppressed but not eliminated as expected for a d-wave
gap, and transfer much of their weight to the resonance region.
The spin susceptibility, which is confined to nearly
antiferromagnetic momenta, shows no evidence for an energy gap in
the normal phase, and we suggest that experiments in which a
pseudogap suppression is seen sample a range of lower momenta.
However we find that the resonance and incommensurate response
make their appearance below a temperature as high as T$^{*}$$\sim$
2$T_{c}$ comparable with the pseudogap temperature. The fraction
of the low-temperature resonance weight that develops before
entering the superconducting phase is surprisingly large.  We
suggest that this indicates that substantial superconducting
pairing fluctuations occur in the normal phase of underdoped
YBCO$_{6.5}$.


\begin{acknowledgments}

We have benefited from discussions with I. Affleck, P. Bourges,
J.P. Carbotte, S. Chakravarty, A.V. Chubukov, R. A. Cowley, R.
Coldea, P. Dai, S. Kivelson, R. Laughlin, K. Levin, P.A. Lee, M.
R. Norman, D. Pines, S. Sachdev, B. Statt, L. Taillefer and J.
Tranquada.  We are grateful to R. L. Donaberger, L. E. McEwan, A.
Cull and M. M. Potter for superb technical assistance at Chalk
River. The work at the University of Toronto and the University of
British Columbia was supported by the Natural Sciences and
Engineering Research Council (NSERC) of Canada.  C. Stock
acknowledges a GSSSP supplement from the National Research Council
(NRC) of Canada.
\end{acknowledgments}

\section{Appendix A: Absolute Units and Calibration of Spectrometer}

    To compare our data with that of other groups and to theory we have put
our measurements of $\chi''$(\textbf{Q},$\omega$) on an absolute
scale by comparing the measured magnetic intensity to the
integrated intensity of a transverse acoustic phonon near the (0 0
6) Bragg peak, e.g. with \textbf{Q} =(0.15, 0.15, 6).  The use of
a phonon from the sample provides a good internal calibration
which is independent of errors resulting from impurities or
domains which may be an issue with the use of a vanadium standard.

    For  fixed final energy and counting time determined by
 the fixed counts in an incident beam monitor whose efficiency varied
with incident wave vector as 1/k$_{0}$, the measured intensity, I,
is directly proportional to the magnetic or phonon cross-section,

\begin{eqnarray}
{I(\bf Q,\omega)}\propto S(\bf Q, \omega).
\end{eqnarray}

The constant of proportionality (\textit{A}) can be determined
from the measured integrated intensity, $I(\bf Q)$=$\int$
d$\omega$ I($\bf{Q}$, \ $\omega$), of an acoustic phonon. In the
long wavelength limit

\begin{eqnarray}
I(\bf Q) &= A \left({\hbar \over 2 \omega_p}\right)
\left[{1+n(\omega_p)}\right]{|F_N|^2} {Q^2 \cos^2(\beta) \over
M}{e^{-2W}},
\end{eqnarray}

\noindent where \textit{M} is the known mass of the unit cell,
$e^{-2W} \sim 1$ is the Debye-Waller factor, [$1+n(\omega)$] is
the Bose factor, $|F_N|^2$ is the static structure factor of the
Bragg reflection nearest to where the acoustic phonon is measured,
and $\beta$ is the angle between $\bf Q$ and the phonon
eigenvector. Inserting the measured phonon frequency allows a
direct measurement of the calibration factor \textit{A} and
therefore allows any measured intensity to be put on an absolute
scale.

    For the magnetic scattering we relate the measured scattering
to the magnetic correlation function
by~\cite{Buyers_Holden85:Handbook}

\begin{eqnarray}
{S(\bf Q, \omega)}= {g^{2}} {f^{2}(\bf Q)} {B^{2}(\bf Q) }
\sum_{\alpha \beta} {\left(\delta_{\alpha
\beta}-{\widehat{Q}_{\alpha} \widehat{Q}_{\beta} } \right)} \\
\nonumber S_{\alpha \beta}(\bf Q, \omega).
\end{eqnarray}

\noindent The correlation function is related by the fluctuation
dissipation theorem to the imaginary part of the spin
susceptibility $\chi(\bf{Q},\omega)$ by

\begin{eqnarray}
{S_{\alpha \beta}(\bf Q, \omega)}= \pi^{-1} [n(\omega)+1]
{{\chi''_{\alpha \beta}(\bf Q, \ \omega)} \over {g^{2} \mu_B^2}}.
\end{eqnarray}

Since the paramagnetic scattering is isotropic in spin,
$\chi''=\chi''_{xx}=\chi''_{yy}=\chi''_{zz}$, we can extract
$\chi''(\bf {Q},\omega)$ from the measured intensity, I, through
the equation

\begin{eqnarray}
{I(\bf Q,\omega)}= A {(\gamma r_{o})^2  \over 4} {f^2(\bf
Q)}{B^2(Q_{z})}{e^{-2W}} \nonumber \\{[1+n(\omega)] \over \pi
\mu_B^2}({2\chi''(\bf{Q}},{\omega)}).
\end{eqnarray}

\begin{figure}[t]
\includegraphics [width=85mm] {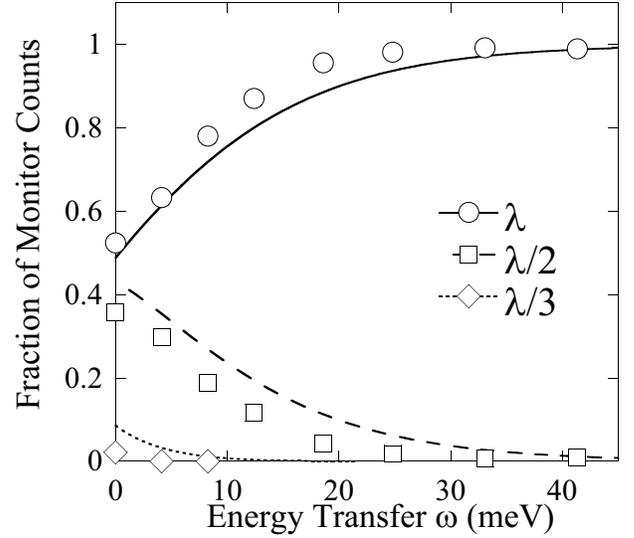}

\caption{The fraction of neutrons hitting the diffracted beam
monitor is plotted as a function of energy transfer for
$\lambda$/n = $\lambda$/1, $\lambda$/2, and $\lambda$/3.  The
final energy is taken to be 14.5 meV where then energy transfer is
defined as $\hbar \omega$=E$_{i}$-E$_{f}$.  The solid curves are
calculations using a Maxwellian distribution and a moderator
temperature $k_{B}T$=28.8 meV.} \label{higher_order_correct}
\end{figure}

\noindent The factor 2 comes from the orientation factor and
$\chi''$ denotes the susceptibility per formula unit here and in
the body of paper. We have chosen to use the anisotropic form
factor $f^2(\bf Q)$ over the isotropic form~\cite{Watson61:14}
because the anisotropic form has been carefully verified in the
ordered antiferromagnetic system and in the Al doped YBCO
system.~\cite{Shamoto93:48, Brecht95:52} For practical
cross-section calculations $(\gamma r_{o})^2  \over 4$ is 73
mbarns sr$^{-1}$. The paramagnetic description for the spin
scattering follows from our polarized measurements and from our
variation of the orientation factor in different Brillouin zones.
These show no preferred polarization of the fluctuations. The
factor $B({Q_{z}})$ is the bilayer structure factor equal to
$sin(Q_{z}d/2)$, where $d$ is the intra-bilayer spacing. In the
normalization to absolute units of the spin fluctuations below
$\sim$ 40 meV energy transfers we have only considered the
acoustic, or odd, fluctuations to the susceptibility since these
predominate at the momenta studied. In the notation of Fong
\textit{et al.} we have calculated $\chi_{odd}''$($\bf{Q}$,
$\omega$) only.

The definitions of the magnetic cross-section used here are
identical to those used in the phonon calibration by Fong
\textit{et al.}~\cite{Fong00:61}. But, the definition of the q and
$\omega$ integrated value of $\chi''$ used by Fong \textit{et al.}
is larger by a factor of $\pi$ from the definition used in this
paper and that of Dai \textit{et al.} We have therefore removed
this factor $\pi$ from the integral quoted by Fong \textit{et al.}
for the comparisons made here.~\cite{Bourgespriv} The vanadium
calibrations conducted by Dai \textit{et al.}~\cite{Dai99:284} use
the isotropic form factor but otherwise the same equations were
used~\cite{Hayden98:241} as presented here.  This allows a direct
comparison between different experiments.~\cite{Absolute}

\section{Appendix B: Correction for Higher Order Contamination of Monitor}

    The low-efficiency fission monitor in the incident beam
only approximately removes the factor 1/$k_{0}$ from the measured
cross-section. As the incident energy decreases, the monitor rate
is perturbed because it senses an increasing fraction of
higher-order neutrons reflected by the monochromator out of the
reactor spectrum . Thus low-energy transfer scattering would be
underestimated if no correction were made. To understand the
relative intensities as a function of energy transfer, a complete
characterization and correction for this contamination is
essential.

    We determined the relative weight of $\lambda$/2 and
$\lambda$/3 neutrons over the energy range of our experiments from
the relative intensities of aluminum powder Bragg peaks in
two-axis mode.  The relative flux of each component of the
incident beam can be uniquely calculated from the following
formula~\cite{Squires},

\begin{eqnarray}
{I\left(2\theta\right)} &\propto& {\Phi(\lambda)} {\lambda^3 \over
\sin(2\theta) \sin(\theta)} {\sum\limits_{\bf Q}} {|F_N(\bf Q)|^2}
\end{eqnarray}

\noindent where $\Phi$ is the incident flux, $|F_N(\bf Q)|^2$ is
the nuclear structure factors and the sum is over constant $|\bf
Q|$.

    By correcting for the 1/$k_{0}$ monitor efficiency the
relative fraction of monitor counts in the beam from $\lambda$,
$\lambda/2$, and $\lambda/3$ orders were calculated as shown in
Fig. \ref{higher_order_correct}. We also computed the fractions
for a Maxwellian distribution of neutrons, a moderator temperature
of $kT=$28.8 meV (60$^{\circ}$ C), and for these three components
only. The resultant fraction of each order is~\cite{Bacon}

\begin{eqnarray}
{F(n^2, E_i)}={(n^2)e^{-(n^2 E_i)/kT}\over e^{-E_i/kT}+4e^{-4E_i/kT}+9e^{-9E_i/kT}}.
\end{eqnarray}

\noindent As shown for $n=$1, 2, and 3 in Fig.
\ref{higher_order_correct}  this simple model provides a
reasonable description of the data given that we have not
corrected for the energy dependent reflectivity of the
monochromator. As can be seen from Fig. \ref{higher_order_correct}
the correction factor required for low energies is substantial.
Inclusion of this factor is essential for obtaining an accurate
form for the spectral distribution, as we have done from 0 to 43
meV. The correction becomes particularly important when discussing
low-energy excitations where it amounts to a factor two.   Neglect
of the factor would make it more difficult to observe low-energy
scattering and might lead to an inference that a normal state spin
gap was present.


\end{document}